\documentclass{aa}

\usepackage[varg]{txfonts}
\usepackage{graphicx}
\usepackage{amssymb}
\usepackage{amsmath}
\usepackage{color}
\usepackage{natbib}
\usepackage{hyperref}

\bibpunct{(}{)}{;}{a}{}{,}

\title{Orbital variability of polarized X-ray radiation reflected from a companion star in X-ray binaries}

\titlerunning{X-ray reflection from a companion star}

\author{Varpu Ahlberg\inst{1}, 
Vadim Kravtsov\inst{1}, Juri Poutanen\inst{1}}

\authorrunning{V. Ahlberg et al.}

\institute{Department of Physics and Astronomy, FI-20014 University of Turku, Finland\\ \email{varpu.a.ahlberg@utu.fi} 
}

\date{Received XXXX / Accepted XXXX}

\begin{document}

\abstract{The reflection of X-ray radiation produced near a compact object from its stellar companion contributes to the orbital variability of polarization in X-ray binaries.
The X-rays are reflected mainly via Thomson scattering resulting in a high polarization. 
The orbital variability of the polarization strongly depends on the inclination and the orbital parameters allowing us to constrain them.
To explore this phenomenon, we present analytical single-scattering models for the polarized reflection.
We find that while diluted by the direct emission, the reflection can produce a polarization degree of about 1\% in the case of a large reflection albedo.
We fitted the orbital variations of the X-ray polarization observed by the Imaging X-ray Polarimetry Explorer from an accreting weakly magnetized neutron star `clocked burster' \mbox{GS~1826$-$238} and found that the amplitude of the variations is too large to be primarily caused by the companion star.
The polarized reflection is more significant if the compact object is obscured from the observer, and thus it should be more easily observable in certain high-inclination targets.}


\maketitle


\section{Introduction}

X-ray binaries (XRBs) comprise a compact object, a black hole or a neutron star, that accretes matter from a stellar companion. 
A fraction of the X-ray emission produced in the vicinity of a compact object is reflected from the companion star. 
This fraction is defined mostly by the ratio of the Roche lobe size to the separation, which is a function of the mass ratio \citep{Eggleton83,Frank02}. 
Soft X-rays are mostly absorbed in the stellar atmosphere, but the harder X-rays are reflected through electron scattering, which incurs a high linear polarization on the reflected light.
The orbital motion of the companion leads to a variation of the X-ray polarization degree and angle \citep{GS74}.
In principle, this polarization may be used to constrain the orbital parameters of the XRBs.

Optical polarization has been used for decades as a tool to study orbital parameters, inclination, and orientation on the sky \citep{Brown78} in massive binary stars \citep{Berdyugin16,Berdyugin18,AbdulQadir23}, exoplanets \citep{Berdyugina11,MB12}, gamma-ray binaries \citep{Kravtsov20}, as well as X-ray binaries \citep{Kemp78,Dolan89CygX1,Dolan89A0620,Kravtsov23}. 
In the X-rays, polarimetry in the 2--8 keV band has recently been made possible with the launch in December 2021 of the Imaging X-ray Polarimetry Explorer (IXPE) \citep{IXPE22}.
The contribution of the companion star reflection is small, but it may be marginally detectable within the accuracy of IXPE. 
For example, low-mass X-ray binary \mbox{GS~1826$-$238} exhibits weak but detectable orbital polarization variations \citep{Rankin24}, which have been described using an optically thin electron-scattering model \citep{Brown78,Kravtsov20}.
IXPE did not detect any orbital polarization variations in \mbox{Cyg X-1}, so the reflected component may be too faint to be observable \citep{Krawczynki21}.
Moreover, resolving the orbital polarization is difficult if the orbital period is long.
The IXPE observations of \mbox{LMC X-1} hinted at a variability of the polarization with the orbital period, but it was observed only for two and a half periods \citep{Podgorny23}.
\mbox{Cyg X-3} has high polarization varying with the orbital phase, but produced by processes other than stellar reflection \citep{Veledina23}.

The fraction of the incident light reflected by the star depends on the energy of the photons and the composition of the stellar atmosphere.
For an atmosphere of cosmic abundances with low ionization, the photoionization cross-section $\sigma_\mathrm{ph}$ is approximately equal to the Thomson scattering cross-section $\sigma_\mathrm{T}$ at 8 keV and reduces as $\propto E^{-3}$ with increasing energy.
Below this threshold, most of the incoming radiation is absorbed and reprocessed to lower energies.
At energies greater than 8 keV the electron scattering cross-section dominates over absorption and most of the X-rays will be reflected \citep{BS74}. 
If the reflected photons undergo only one scattering, their polarization degree (PD) is \citep{Cha60}
\begin{equation} \label{eq:poldeg}
P = \frac{1-\mu^2}{1+\mu^2},
\end{equation} 
where $\mu$ is the cosine scattering angle.
Single-scattered light can therefore be fully polarized, but further scatterings reduce the PD.
The number of scatterings depends on the single-scattering albedo, which is the ratio of the scattering and total absorption coefficients, $\lambda = \alpha_\mathrm{T}/(\alpha_\mathrm{T} + \alpha_\mathrm{ph})$. 
Thus in the standard X-ray band 2--10 keV the reflection is well approximated with single scattering.
The scattering albedo in the soft X-rays is small for a normal stellar atmosphere, although it may be enhanced due to the effects of irradiation.
Specifically, some of the energy absorbed by the companion is transformed into evaporative winds near the surface layers of the atmosphere \citep{Blondin94}.
The outflowing gas is hot and highly ionized, and therefore absorption is negligible compared to scattering.
The gas would have a significant Thomson optical depth, and its reflection albedo remains nearly constant in the soft X-rays \citep{BS74}.
Different XRBs likely have different albedos, so the amplitude of the orbital polarization may vary from target to target.

In this paper, we present analytical single-scattering models for the X-ray stellar reflection in XRBs.
In Sect.~\ref{sec:model} we detail the geometry and polarization basis of our models, and in Sect.~\ref{sec:reflection} we describe the reflected flux under different approximations.
We move on to study how the models behave with different parameters in Sect.~\ref{sec:results}.
We then apply the model to observations of an accreting neutron star \mbox{GS~1826$-$238} in Sect.~\ref{sec:applications}, and discuss the results in Sect.~\ref{sec:discussion}.

\begin{figure} 
\begin{center}
\includegraphics[width=\linewidth]{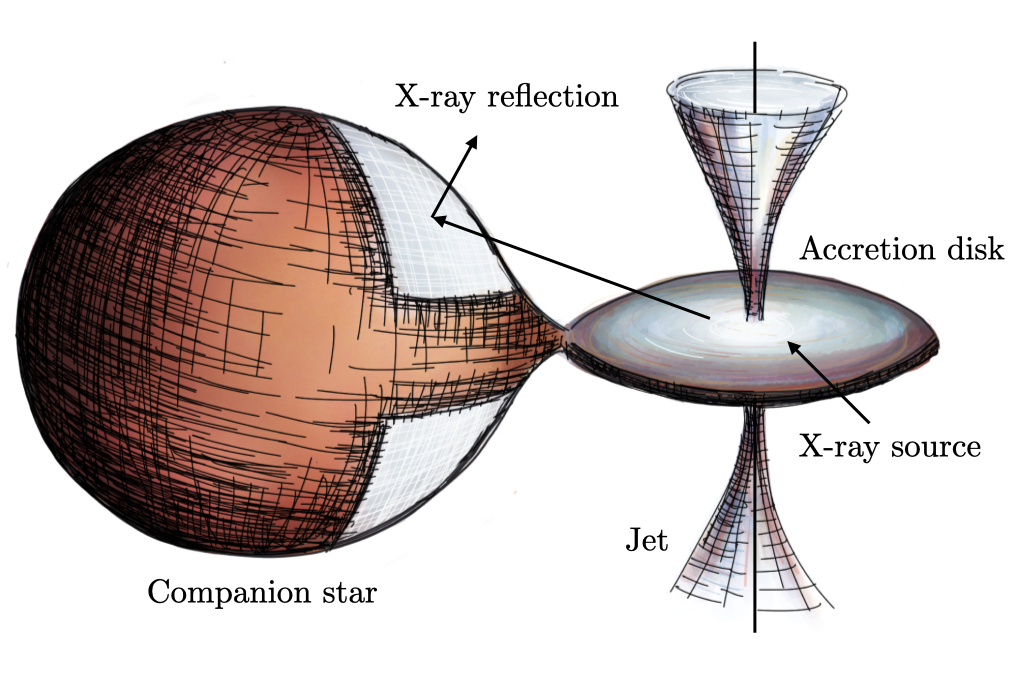}
\end{center}
\caption{Illustration of how X-rays emitted near the compact object are reflected from the binary companion in XRBs.}  
\label{fig:illustration}
\end{figure}

\section{Model} \label{sec:model}

\subsection{Geometry}

\begin{figure} 
\begin{center}
\includegraphics[width=0.75\linewidth]{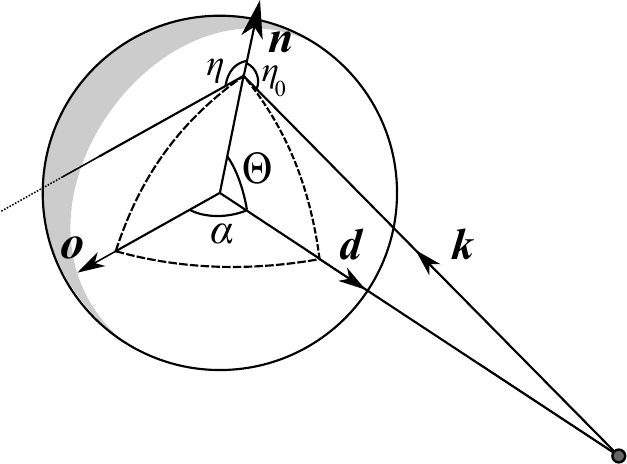}
\end{center}
\caption{Geometry of the reflection model.
The light emitted by a point-like compact object is intercepted by the binary companion and a fraction is reflected toward the observer along vector $\vec{\hat o}$.
\label{fig:binary_geometry}} 
\end{figure}

We considered an X-ray binary containing a point-like compact object
(see Fig.~\ref{fig:illustration} for illustration). 
The compact object emission is assumed to be unpolarized and isotropic.
We first modeled the reflection geometry as a spherical companion of radius $r$ at a binary separation of $d$, as depicted in Fig.~\ref{fig:binary_geometry}.
In the case of the Roche lobe overflow the shape of the star deviates from a sphere, and the shadow of the accretion disk covers the equator.
We did not model the shadowing here, but we make calculations for the Roche lobe in Sect.~\ref{section:roche}.

We chose a coordinate system with the origin coinciding with the companion star and the z-axis aligned with the orbital axis  $\vec{\hat\Omega}$ = (0,0,1). 
The unit vector pointing from the center of the companion star toward the compact object lies on the x-axis: 
\begin{equation}
    \vec{\hat d} = (1, 0, 0). 
\end{equation}
In these coordinates, the direction toward the observer rotates clockwise as a function of the orbital phase angle $\varphi$ (shifted true anomaly):
\begin{equation}
    \vec{\hat o} = (- \sin i \cos \varphi , \sin i \sin \varphi , \cos i),
\end{equation}
where $i$ is the inclination of the observer to the orbital axis.
With this definition, the star is between the observer and the compact object when $\varphi = 0$.
The binary separation varies as
\begin{equation}
    d(\varphi) = \frac{a(1 - e^2)}{1 + e \cos(\varphi - \omega)},
\end{equation}
where $a$ is the semi-major axis of the system, $e$ its eccentricity, and $\omega$ is the phase of the periastron.
The cosine of the phase angle (i.e. the angle between the observer direction and the vector pointing from the center of the companion star toward the compact object) is
\begin{equation}
    \cos \alpha = \vec{\hat d} \cdot \vec{\hat o} = - \sin i \cos \varphi.
\end{equation}
The stellar surface normal can be written as
\begin{equation}
    \vec{\hat n} = (\sin \theta \cos \phi, \sin \theta \sin \phi, \cos \theta),
\end{equation}
where $\theta$ and $\phi$ are the co-latitude and azimuthal angle.
The cosine angle between the reflected photons propagating toward the observer and the surface normal is
\begin{equation}
    \eta = \vec{\hat o} \cdot \vec{\hat n} = \cos i \cos \theta - \sin i  \sin \theta  \cos (\phi + \varphi).
\end{equation}
The direction of the incident light, $\vec{\hat k}$, can be calculated as a linear combination of vectors $\vec{\hat d}$ and $\vec{\hat n}$. 
First, we define the angle between the surface normal and the orbital vector as
\begin{equation}
    \cos \Theta = \vec{\hat n} \cdot \vec{\hat d} 
    = \sin \theta \cos \phi .
\end{equation}
Using the law of cosines, the distance between the compact object and the point on the surface is
\begin{equation}
    k^2 = d^2 + r^2 - 2rd \cos \Theta . 
\end{equation}
Vector $\vec{\hat k}$ can be expressed as
\begin{equation}
    \vec{\hat k} =  \dfrac{r}{k} \vec{\hat n} - \dfrac{d}{k} \vec{\hat d},
\end{equation} 
which yields the cosine angle between the incident light and the surface normal
\begin{equation}
 \eta_0 = - \vec{\hat n} \cdot \vec{\hat k}= \dfrac{d \cos \Theta - r}{k},
\end{equation}
and the cosine of the scattering angle
\begin{equation}
    \mu = \vec{\hat k} \cdot \vec{\hat o} = 
    \dfrac{r\, \eta - d \cos \alpha }{k}.
\end{equation}

\subsection{Visibility conditions}

The visibility of the reflected light depends on the overlap between the stellar area visible to the observer and the area illuminated by the compact object.
Firstly, the element of the stellar surface has to be visible from the compact object: 
\begin{equation}
    \eta_0 > 0,
\end{equation}
which is satisfied when $\cos \Theta > r/d$.
This limits the visible area to
\begin{align}
    \arcsin \left( \frac{r}{d} \right) &< \theta < \pi - \arcsin \left( \frac{r}{d} \right), \\
    -\arccos \left( \frac{r}{d \sin \theta} \right) &< \phi < \arccos \left( \frac{r}{d \sin \theta} \right).
\end{align}
Secondly, the surface element must be visible to the observer as well:
\begin{equation}
    \eta > 0.
\end{equation}
The illuminated surface is completely invisible for $\alpha$ close to $\pi$, when 
\begin{equation} \label{eq:invisible}
    \cos \alpha < 0 \ \mbox{and} \ \sin \alpha < r/d .
\end{equation}
The visible range of angles for the observer is
\begin{align}
    i - \pi/2 &< \theta < i + \pi/2, \\
     \arccos (\cot i \cot \theta) &< \phi + \varphi < 2 \pi - \arccos (\cot i \cot \theta).
\end{align}
If $\theta < \pi/2 - i$, the surface is visible to the observer for all $\phi$.
The combination of the two visibility conditions can be complicated, as the visible ranges of $\phi$ can overlap in two separate intervals.

\subsection{Polarized reflection} \label{sec:polreflection}

Linear polarization of the reflected radiation is fully described by the Stokes parameters $I$, $Q$, and $U$. 
The PD is $P = \sqrt{Q^2 + U^2}/I$ and the normalized Stokes parameters $q = Q/I$ and $u = U/I$ can be written as
\begin{equation}
q =   P \cos(2\chi), \qquad 
u =   P \sin(2\chi) ,
\end{equation} 
where $\chi \equiv (1/2)\arctan(U/Q)$ is the polarization angle (PA).
The angle depends on the choice of polarization basis, which we defined by the projection of the orbital axis on the plane of the sky:
\begin{align} \label{eq:basis_main}
\vec{\hat e}_1  &= \frac{\vec{\hat \Omega} - \cos i \ \vec{\hat o} }{\sin i } = ( \cos i \cos \varphi , - \cos i \sin \varphi , \sin i), \\
\vec{\hat e}_2 &= 
\frac{\vec{\hat o} \times \vec{\hat \Omega} }{\sin i } = (\sin \varphi, \cos \varphi, 0). 
\end{align} 
The scattering plane can be expressed using the polarization pseudo-vector:
\begin{equation}
  \vec{\hat p} = \frac{\vec{\hat o}\times\vec{\hat k}}{|\vec{\hat o}\times\vec{\hat k}|} .
\end{equation}
The PA is the angle between the polarization vector and the basis:
\begin{align}
   \cos\chi &= \vec{\hat e}_1 \cdot \vec{\hat p}
   = \frac{d \sin \varphi - r \sin \theta \sin (\phi + \varphi)}{k\sqrt{1-\mu^2}}, \\  
  \sin\chi &= \vec{\hat e}_2 \cdot \vec{\hat p}  \\ 
   &= \frac{r[\sin i \cos \theta + \cos i \sin \theta \cos (\phi + \varphi)] - d \cos i \cos \varphi}{k\sqrt{1-\mu^2}} \nonumber .
\end{align}
The PD after single scattering is determined by Eq. \eqref{eq:poldeg}.
If the compact object's intrinsic emission has a small PD, it does not considerably change the polarization of the reflection emission. 
Thus, a constant term corresponding to the intrinsic polarization can be added to the Stokes parameters to model its contribution.
   
\begin{figure} 
\begin{center}
\includegraphics[width=0.95\linewidth]{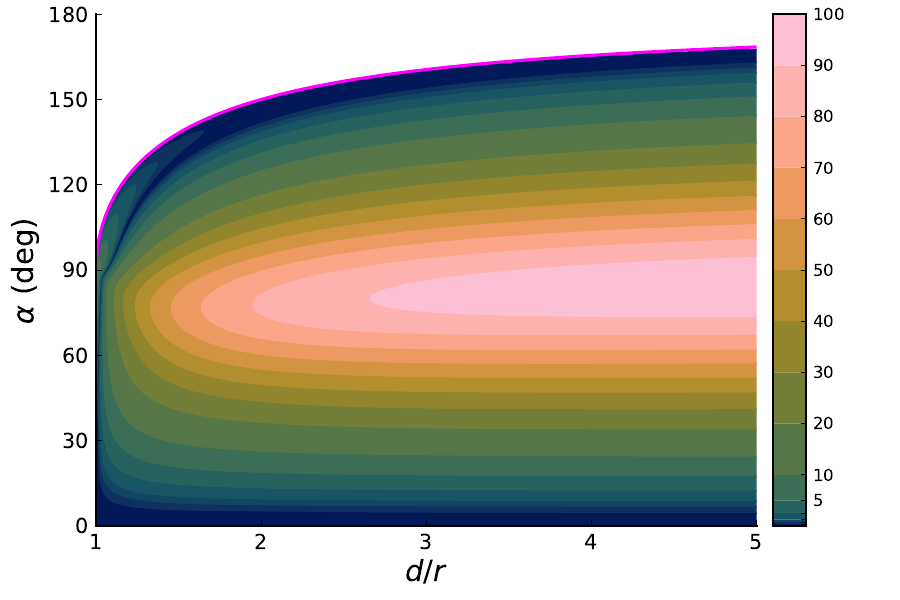}
\end{center}
\caption{PD (in \%) of light reflected from the surface of a spherical star as a function of the phase angle $\alpha$ and separation parameter $d/r$. The magenta line represents the eclipse limit given by Eq.~\eqref{eq:invisible}, above which the reflected flux is zero (white area). To highlight the behavior near eclipses, each contour below 10\% is half the PD of the previous one, down to 0.375\%.
\label{fig:poldeg}} 
\end{figure}

\subsection{Scattering from a star filling its Roche lobe} \label{section:roche}

In some XRBs, the companion star loses its mass through Roche lobe overflow.
The Roche lobe is described as the equipotential surface that includes the first Lagrange point (L1), which is a point along the x-axis where the gradient of the gravitational potential is zero.
The dimensionless gravitational potential under synchronous rotation is \citep{Leahy2015}
\begin{align}
\label{eq:potential}
    \psi &= \frac{1}{\rho}  + q_\mathrm{m} \left( \frac{1}{\sqrt{1 - 2\rho \sin \theta \cos \phi + \rho^2}} -\rho \sin \theta \cos \phi \right) \nonumber \\ 
&+ \frac{q_\mathrm{m}+1}{2} \rho^2 \sin^2 \theta,
\end{align}
where $\rho$ is the radial distance from the center of the star in units of binary separation and $q_\mathrm{m} = m_\mathrm{x}/m_\mathrm{c}$ is the ratio of the compact object mass $m_\mathrm{x}$ and the companion mass $m_\mathrm{c}$.
Expressed using Cartesian coordinates, the gradient of the potential is 
\begin{align}
    \frac{\mathrm{d} \psi}{\mathrm{d} x} &= \frac{x}{\rho^3} - q_\mathrm{m} \left( \frac{1-x}{(1 - 2x + \rho^2)^{3/2}} - 1\right) - (q_\mathrm{m}+1)x, \\
    \frac{\mathrm{d} \psi}{\mathrm{d} y} &= \frac{y}{\rho^3} + q_\mathrm{m} \frac{y}{(1 - 2x + \rho^2)^{3/2}} - (  q_\mathrm{m}+1)y, \\
    \frac{\mathrm{d} \psi}{\mathrm{d} z} &= \frac{z}{\rho^3} + q_\mathrm{m} \frac{z}{(1 - 2x + \rho^2)^{3/2}},
\end{align}
where $\rho = \sqrt{x^2 + y^2 + z^2}$.
The position of the L1 point and the value of the potential there can be found from the condition $\mathrm{d} \psi/\mathrm{d} x=0$ at $y=z=0$ as a function of $q_\mathrm{m}$, although it is a fifth-order polynomial and thus requires some numerical root-finding method.
Finding the shape of the Roche lobe $\rho(\theta,\phi)$ where the potential is equal to that at L1 has to be done numerically as well.
The surface normal of the Roche lobe, $\vec{\hat n}^*$, is the unit vector of the gradient along the equipotential surface.
Similar to the spherical star, the cosine scattering angles are $\eta = \vec{\hat o} \cdot \vec{\hat n}^*$, $\eta_0 = -\vec{\hat k} \cdot \vec{\hat n}^*$, and $\mu = \vec{\hat k} \cdot \vec{\hat o}$.
The visibility conditions do not have a simple analytical form due to the complexity of this geometry.

\section{Reflected flux from a stellar surface} \label{sec:reflection}

\subsection{Exact calculations}

The radiative transfer equation for a plane-parallel atmosphere with absorption and Thomson scattering is \citep{Cha60}
\begin{equation}
  \eta \frac{\mathrm{d}\tilde{I}(\tau,\eta,\Phi)}{\mathrm{d}\tau}
  =  \frac{1}{\lambda} \tilde{I}(\tau,\eta,\Phi)   
  - \tilde{S}(\tau,\eta,\Phi) , 
\end{equation}
where $\lambda$ is the single-scattering albedo, $\eta$ is the cosine of the zenith angle, $\Phi$ is the azimuthal angle relative to the scattering plane, and $\mathrm{d}\tau=-\sigma_\mathrm{T}n_\mathrm{e} \mathrm{d}z$ the vertical Thomson optical depth.
Using the source function for Thomson/Rayleigh scattering and assuming unpolarized incident light, the Stokes vector of the single-scattered radiation is (see \citealt{Veledina23} and p.146 of \citealt{Cha60})
\begin{equation}
  \tilde{I}_1(\eta,\Phi)  = 
  \frac{3}{16\pi} I_0 \ \lambda  \ (1+\mu^2)\ 
  \left( \begin{matrix} 
    1 \\
    P \cos2\chi \\
    P \sin2\chi
  \end{matrix} \right)
  \frac{\eta_0}{\eta+\eta_0} ,
\end{equation}
where $I_0$ is the flux of the incident light.
In the isotropic case, $I_0 = L/(4\pi k^2)$, where $L$ is the luminosity of the compact object.
The total Stokes vector of the reflected light can be obtained by integrating over the surface of the star.
In spherical coordinates, the surface element of a sphere at a constant radius is
\begin{equation} 
\mathrm{d} S= r^2 \sin \theta \,\mathrm{d} \theta \ \mathrm{d} \phi.
\end{equation}
In the case of a nonspherical star, the element is
\begin{equation} 
  \mathrm{d} S = r^2 \sqrt{
    \sin^2 \theta + \left( \frac{\sin \theta}{r} \frac{\mathrm{d} r}
    {\mathrm{d} \theta} \right)^2 + \left( \frac{1}{r} 
    \frac{\mathrm{d} r}{\mathrm{d} \phi} \right)^2
  }  \,\mathrm{d} \theta \, \mathrm{d} \phi.
\end{equation}
For a Roche lobe, we estimated the radius derivatives numerically.
The reflected flux from the projection of a surface element is
\begin{equation} 
\mathrm{d} \tilde{F}_\mathrm{r}= \frac{\eta\ \mathrm{d} S}{D^2}\ \tilde{I}_1 (\theta,\phi) , 
\end{equation}
where $D$ is the distance of the observer.
Thus, the reflected flux (Stokes vector) from a spherical star is
\begin{equation} \label{eq:integral}
  \tilde{F}_\mathrm{r} = F_\star 
  \frac{3\lambda }{16\pi}   
  \int\limits_0^{\pi}  r^2 \sin \theta \,\mathrm{d}\theta 
  \int\limits_0^{2\pi} \frac{1}{k^2} \  (1+\mu^2)\ 
  \left( \begin{matrix}
    1 \\
    P \cos2\chi \\
    P \sin2\chi
  \end{matrix} \right) 
  \frac{\eta \eta_0}{\eta+\eta_0} \mathrm{d}\phi ,
\end{equation}
 where $F_\star = L/(4\pi D^2)$ is the direct flux of the compact object.
This integral can be performed using standard quadrature methods.
We used the visibility conditions to set the integral limits for the spherical star.
For the Roche lobe geometry we integrated over the entire surface, but set the flux to zero when the visibility conditions were not met.
Due to symmetry around the x-axis, the integrated PD of the spherical star reflection depends only on $\alpha$ and $d/r$, as is shown in Fig.~\ref{fig:poldeg}.
However, the asymmetric shape of the Roche lobe makes its PD depend on $i$ and $\varphi$ separately.

The observed Stokes vector is a sum of the reflected component and the direct unpolarized emission:
\begin{equation}
\tilde{F}_\mathrm{tot}=  
 F_\star \left( \begin{matrix} 
1 \\  0 \\  0  \end{matrix} \right)   + \tilde{F}_\mathrm{r} .
\end{equation} 
The direct emission therefore dilutes the observed PD depending on the amount of reflected light.
The observed PD is $P_\mathrm{obs} = F_\mathrm{r} P /({F_\mathrm{r} + F_\star}) $. 
The PA is computed from the $Q$ and $U$ components of the Stokes vector $\tilde{F}_\mathrm{r}$. 

\subsection{Large separation approximation}

The reflected flux can be solved analytically if a very large binary separation is assumed.
For small values of  $r/d$, the direction of incident light is $\vec{\hat k} \approx \vec{\hat d}$.
It follows that $\eta_0 \approx \cos \Theta$ and $\mu \approx -\cos \alpha$. 
The PD becomes
\begin{equation} \label{eq:distant_poldeg}
    P = \frac{1 - \cos^2 \alpha}{1 + \cos^2 \alpha}, 
\end{equation}
and the PA
\begin{align}
    \sin \chi &= -\frac{\cos i \cos \varphi }{\sqrt{1 - \cos^2 \alpha}}  , \label{eq:lambertsine}\\
    \cos \chi &= \frac{\sin \varphi}{\sqrt{1 - \cos^2 \alpha}} \label{eq:lambertcosine}.
\end{align}
Under this approximation, the integral in Eq. \eqref{eq:integral} becomes analytically solvable \citep[p.~192 in][]{Sobolev75}:
\begin{align}
    F_\mathrm{r}  &=  \epsilon F_\star \frac{3\lambda}{8} (1 + \cos^2 \alpha) \Phi_\mathrm{LS}(\alpha), \\
    \epsilon &= \frac{1}{2} \left( 1 - \sqrt{1 - \frac{r^2}{d^2} }\right),\\
    \Phi_\mathrm{LS}(\alpha) &= 1 - \sin \frac{\alpha}{2} \tan \frac{\alpha}{2} \ln \left[\cot \frac{\alpha}{4}\right] ,
\end{align}
where $\epsilon$ is the fraction of the compact object flux intercepted by the star and $\Phi_\mathrm{LS}$ is the Lommel-Seeliger phase function as described in \citet{Russell1916}. 
Assuming $F_\mathrm{r} \ll F_\star$, the normalized Stokes $q$ and $u$ for the diluted reflection is
\begin{align}
    q &= f_0  \left(\sin^2 \varphi  - \cos^2 \varphi \cos^2 i \right) \Phi_\mathrm{LS}(\alpha), \label{eq:normq} \\
    u &= -f_0 \sin 2 \varphi \cos i \ \ \Phi_\mathrm{LS}(\alpha), \label{eq:normu}
\end{align}
where $f_0 = \frac{3}{8} \lambda \epsilon$ is the flux normalization factor.

A different analytical approximation of scattering from a distant spherical object is the Rayleigh-Lambertian reflector.
Using the Lambertian phase function, the reflected flux is \citep{Russell1916}
\begin{align}
    F_\mathrm{r} &= \epsilon F_\star p \Phi_\mathrm{L}(\alpha), \\
    \Phi_\mathrm{L}(\alpha) &= \frac{\sin \alpha + (\pi - \alpha) \cos \alpha}{\pi},
\end{align}
where $p = {2}/{3}$ is the geometrical albedo of a Lambertian disk.
While the Lambertian phase function $\Phi_\mathrm{L}$ assumes isotropic scattering which does not polarize the light, we used the above Thomson scattering formulae to calculate the polarization under this approximation.

\begin{figure} 
\begin{center}
\includegraphics[width=0.95\linewidth]{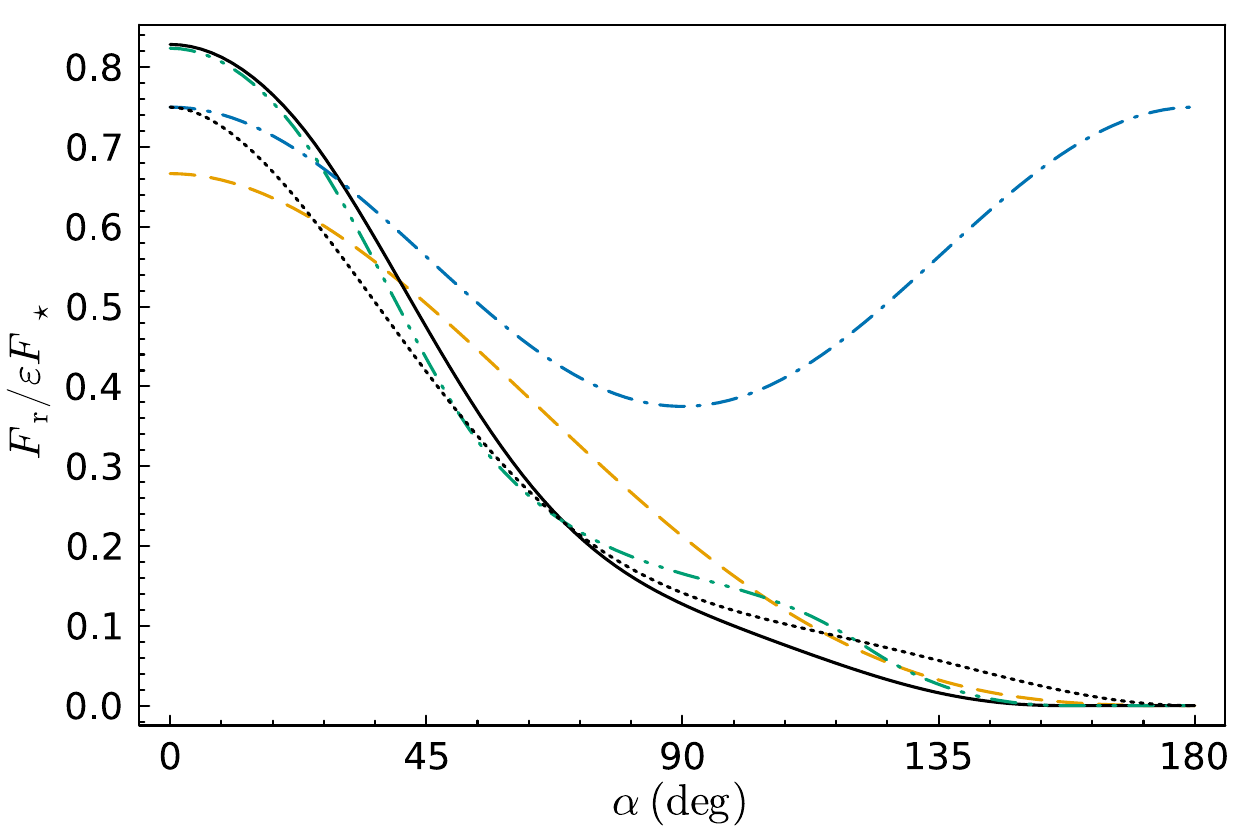}
\end{center}
\caption{Comparison between the phase function of the different models: Thomson scattering spherical star including the exact calculation (solid, black) and the large separation approximation (dotted, black), Lambertian reflector (dashed, orange), Thomson scattering cloud (dash-dotted, blue), and Thomson scattering from a Roche lobe (green, dash-double dotted). The binary separation parameter is $d/r = 2.673$ and the mass ratio $q_\mathrm{m} = 1.0$.
\label{fig:phasecompare}} 
\end{figure}

\begin{figure} 
\begin{center}
\includegraphics[width=0.95\linewidth]{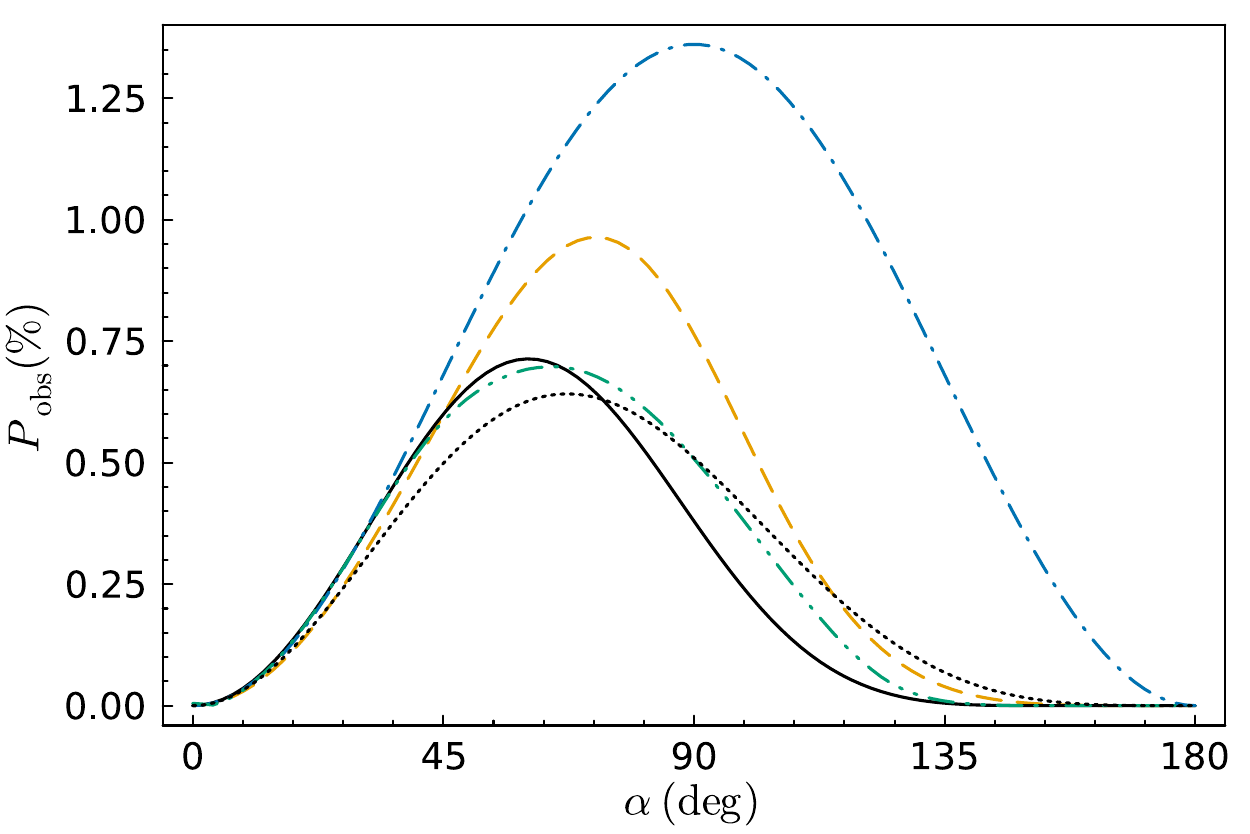}
\end{center}
\caption{Comparison of the diluted PD between the same models as Fig.~\ref{fig:phasecompare}.
\label{fig:poldeg-alpha-m1}} 
\end{figure}

\subsection{Optically thin cloud}

As a point of comparison, we also considered scattering from an orbiting optically thin cloud.
If the cloud is distant, the PD and PA of the reflected radiation are identical to the large separation approximation.
The reflected flux is \citep{Kravtsov20}
\begin{align}
    F_\mathrm{r} &= \epsilon F_\star \frac{3}{8} (1 + \cos^2 \alpha), \\
    \epsilon &= \frac{N_\mathrm{e} \sigma_\mathrm{T}}{4 \pi d^2},
\end{align}
where $N_\mathrm{e}$ is the number of free electrons in the cloud.
The distant cloud does not have a set size or shape, and the fraction of scattered radiation $\epsilon$ is rather determined by the number of scattering electrons.
The density structure and shape of the cloud begin to matter if the cloud is closer to the point source.
Modeling such a cloud is beyond the scope of this work, especially as we did not use it to represent any physical feature in XRBs.
The purpose of the model is to demonstrate the difference between optically thick and optically thin reflecting mediums.

\section{Results} \label{sec:results}

\subsection{Comparison between models}

We calculated the binary companion reflection using Eq. \eqref{eq:integral} for both the spherical and Roche lobe cases and compared it to the large separation approximation, Lambertian reflector, and the optically thin cloud.
Although the Roche lobe reflection depends on both $i$ and $\varphi$ rather than just $\alpha$, we compared it to the other models as a function of $\alpha$ by varying the inclination while keeping the orbital phase angle fixed.
This produces slightly different results than with a fixed inclination, but it does not change the qualitative comparison.
Additionally, as the size of the Roche lobe depends on the mass ratio $q_\mathrm{m}$ rather than $d/r$, we set the size of the spherical star so it corresponded to the radius of the Roche lobe along the y-axis.
We find that this produces results more similar to the spherical star than using the equivalent spherical radius of the lobe's surface area.
For both the cloud and the Roche lobe we set $\epsilon$ equal to that of the spherical models.

Figure \ref{fig:phasecompare} shows the ratio $F_\mathrm{r}/({F_\star \epsilon})$ for all five cases for the conservative limit $\lambda = 1$.
The angular dependence of the reflected flux is similar between the Roche lobe and the spherical star.
Both the large separation approximation and the optically thin cloud reflect ${3}/{4}$ of the incoming flux at maximum, which is the classical result for Thomson scattering.
The spherical and Roche lobe models reach a higher normalized flux because of the different geometry of the reflecting area.
The Rayleigh-Lambert approximation differs significantly from all of the Thomson scattering cases since it uses a different law of reflection.
Unlike all the other models, the optically thin cloud is symmetric around $\alpha = 90\degr$ as the observer always sees the full reflection.

\begin{figure} 
\begin{center}
\includegraphics[width=0.95\linewidth]{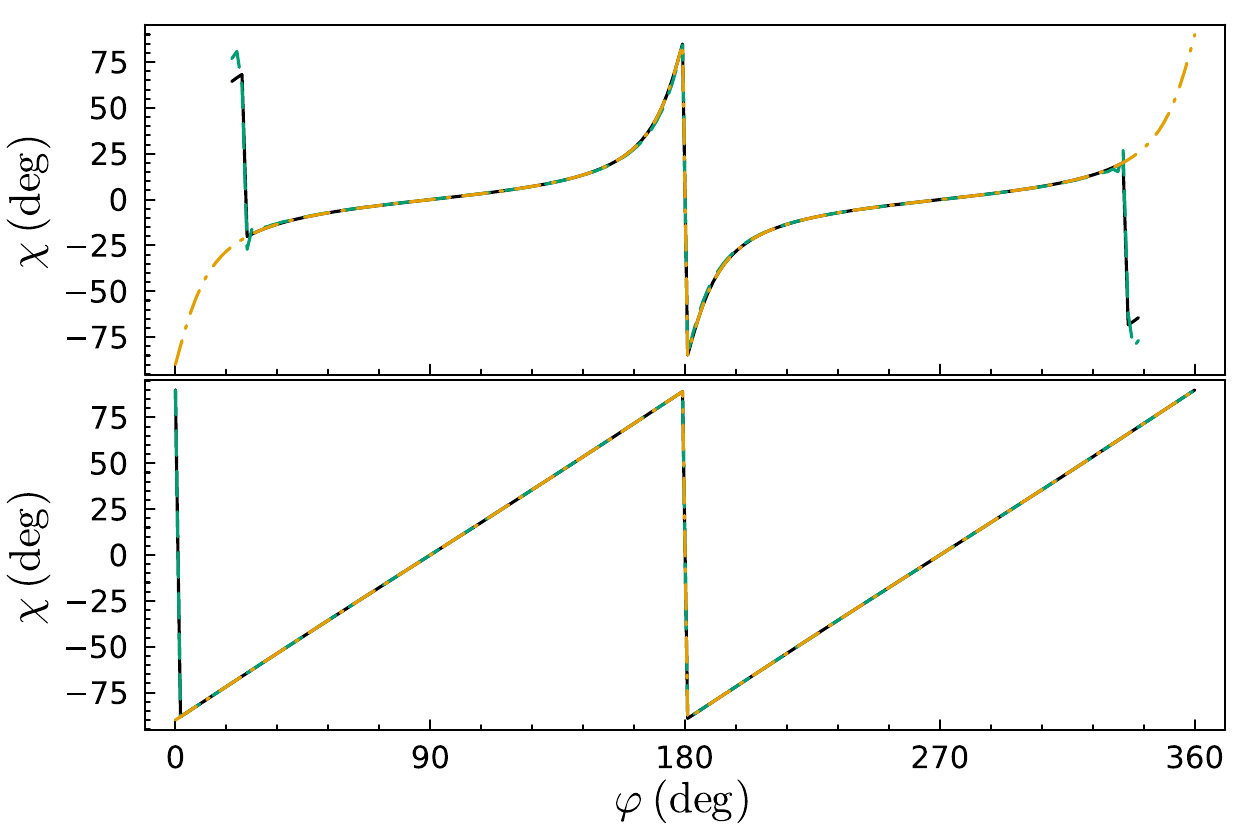}
\end{center}
\caption{PA as a function of orbital phase angle at inclinations of $80\degr$ (top) and $10\degr$ (bottom) for a Thomson scattering spherical star (black, solid) and one filling its Roche lobe (green, dashed), and a distant scatterer (orange, dash-dotted). The orbital separation and mass ratios are the same as in Fig.~\ref{fig:phasecompare} }
\label{fig:polangle} 
\end{figure}

The angular dependence of $P_\mathrm{obs}$ assuming $\lambda = 1$ is shown in Fig.~\ref{fig:poldeg-alpha-m1}.
The models act similarly under $\alpha \lesssim 40\degr$, but diverge as the angle increases.
The spherical and Roche lobe models are the most different at $\alpha \sim 90\degr$, thus the error of assuming a spherical geometry is most significant at low inclinations and at orbital phase angles of $90\degr$ and $270\degr$.
Overall, the spherical star is a good approximation of the Roche lobe as long as $i \gtrsim 45\degr$.
The large separation approximation has a lower maximum PD and is skewed toward higher phase angles as the visibility is less limited.
The Rayleigh-Lambert model is clearly different from the Thomson scattering stars outside of certain orbital phases.
The PD of the optically thin cloud is much higher than the other models and is symmetric like its flux.

\begin{figure*} 
\begin{center}
\includegraphics[width=0.49\linewidth]{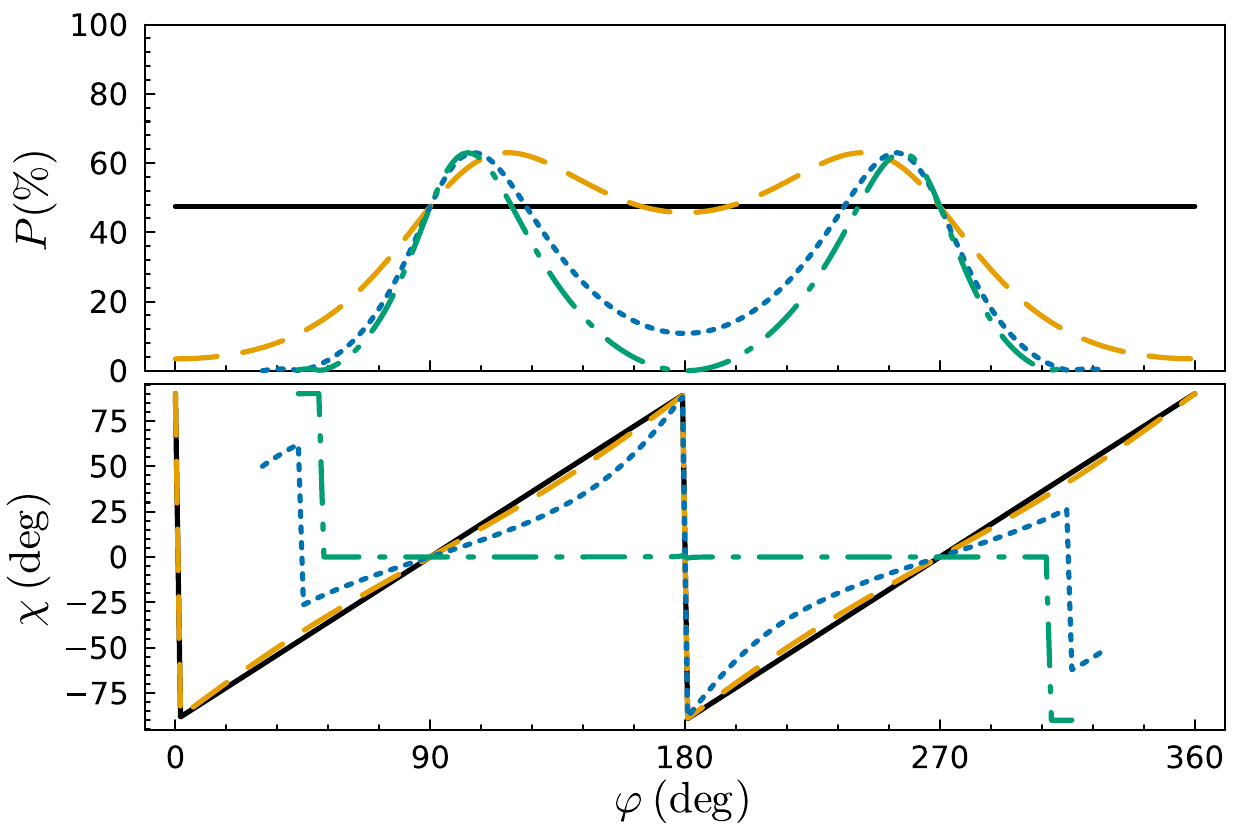}
\includegraphics[width=0.49\linewidth]{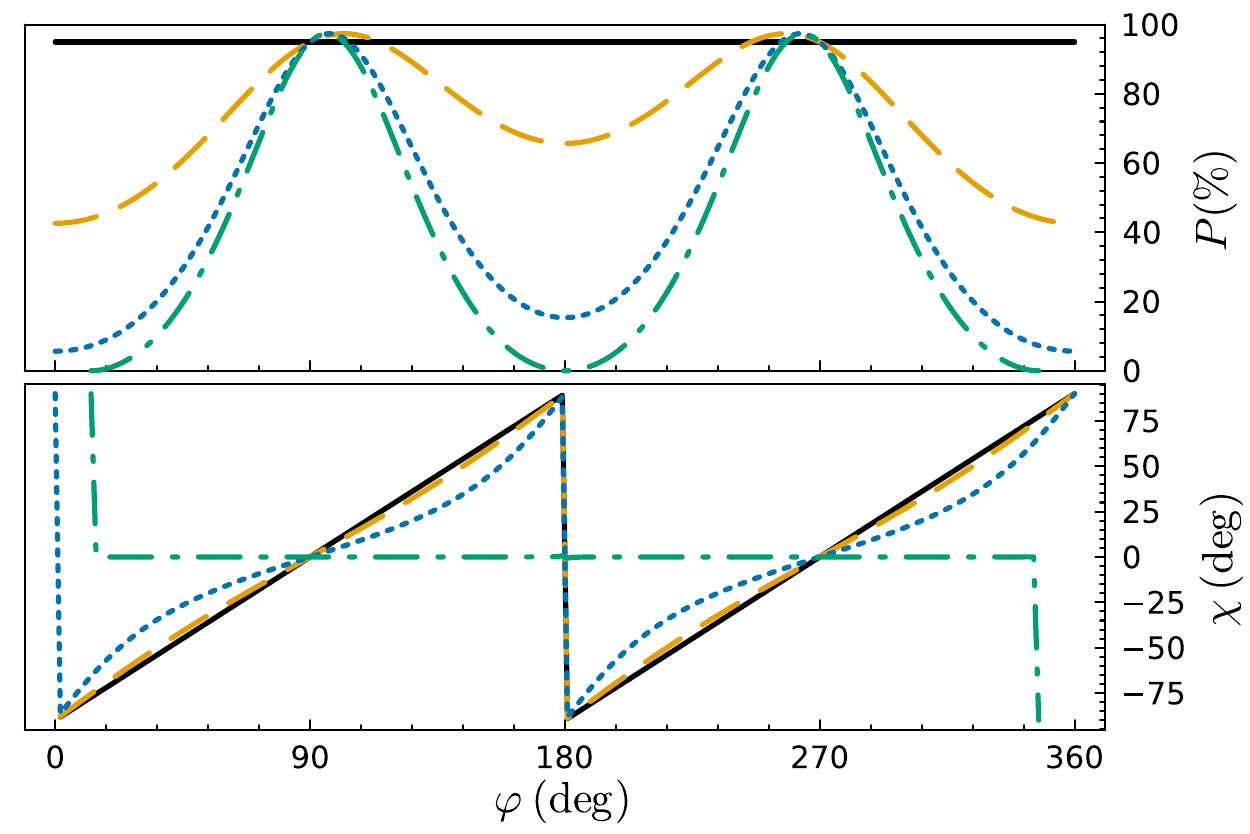}
\end{center}
\caption{Orbital polarization curves of the reflected emission for inclinations of $i=0\degr$ (solid black), $30\degr$ (dashed orange), $60\degr$ (blue dotted), and $90\degr$ (green dash-dotted).
The orbital separation parameter is $d/r = 1.5$ (left) and $5.0$ (right).
\label{fig:phaseplots}} 
\end{figure*}

The PA is nearly the same for each model, as can be seen in Fig.~\ref{fig:polangle}.
Unlike the PD, the PA depends separately on $i$ and $\varphi$ rather than just $\alpha$, so we compared it over one orbital period at two different inclinations.
It undergoes two full rotations each orbit with a different shape depending on the inclination.
The only difference between the models is the presence of eclipses at high inclinations, with $90\degr$ jumps near the eclipse as the PD goes to zero.
The jumps are a consequence of a narrow visible area limiting the scattering angles, making the Stokes $Q$ and $U$ average to zero at some orbital phase.
This can be seen in Fig.~\ref{fig:poldeg} as a narrow contour of zero polarization near the $\sin \alpha < r/d$ eclipse limit.
Besides the eclipse jumps, Eqs.~\eqref{eq:lambertsine} and \eqref{eq:lambertcosine} are an excellent approximation for the PA.

In conclusion, the limited visibility of the stellar surface in close binaries has a significant effect on the reflection.
An optically thin reflector is clearly distinguishable from a star, especially as the polarized flux is much higher than in the other models.
A star filling its Roche lobe can be effectively simplified as a sphere, albeit with some inaccuracy that increases if the inclination is low.
The large separation approximation is the most accurate of the analytical formulae, although it is noticeably different even at moderate separations.
We continue our analysis only for the spherical star because of its mathematical simplicity.

\subsection{Parameter study of the spherical reflector}

We calculated orbital polarization curves of the reflected emission at different inclinations and orbital separations (Fig.~\ref{fig:phaseplots}).
The variability is strongly dependent on the inclination; at $i=0\degr$ the PD remains constant, and at higher inclinations, it has an increasingly double-peaked profile.
Besides the eclipsing behavior, the curve of the PA is entirely determined by the inclination, transforming from a linear profile to a more sinusoidal one as the inclination increases.
At an inclination of exactly $90\degr$, it remains constant over the orbit.

Decreasing the binary separation reduces the PD of the reflected light, because the smaller visible area increases the range of scattering angles.
At $d/r = 5.0$, the difference amounts to only a few percent lower polarization than at large separations, but at $d/r = 1.5$, it is lower by $\sim 50 \%$.
The difference in geometry causes the maximum polarization to occur at smaller phase angles, and therefore at orbital phase angles closer to $180\degr$.
Additionally, the separation determines the length of the eclipses and the phase angle when the PA jumps.
Otherwise, the separation has no noticeable impact on the PA, as evidenced by the accuracy of the large separation approximation.

\begin{figure} 
\begin{center}
\includegraphics[width=0.95\linewidth]{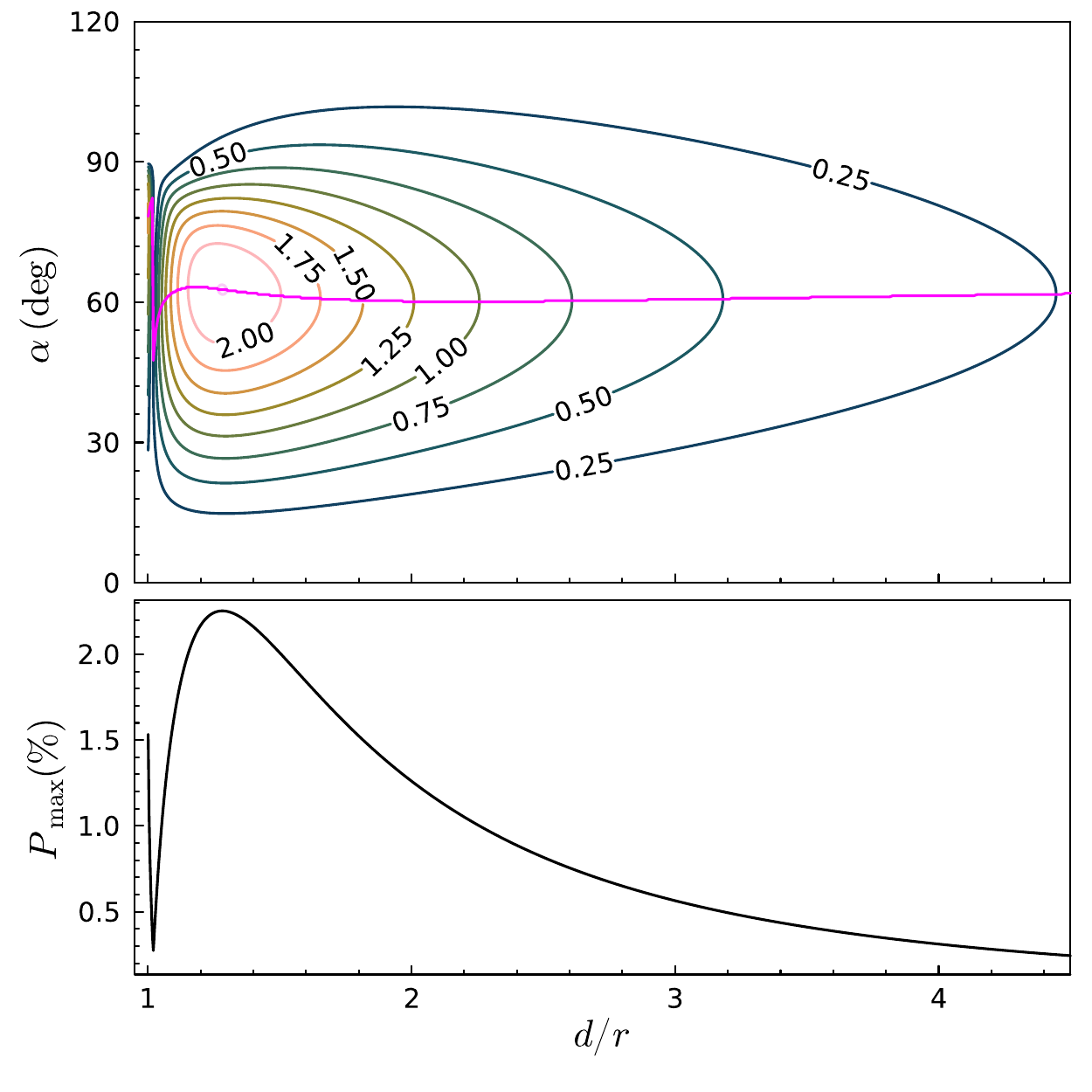}
\end{center}
\caption{Contours of constant PD (in \%) of the total radiation at the plane $d/r-\alpha$ (top). 
The location of the maximum PD for different $d/r$ is shown with a magenta line and its dependence on $d/r$ at the bottom panel. 
\label{fig:maxpol}} 
\end{figure}

Dilution of the reflected radiation by the direct emission reduces the PD of the observed light. 
As we only consider single scattering, the albedo $\lambda$ acts as a simple coefficient for the amount of reflected flux.
While at $\lambda \sim 1$ the light realistically undergoes multiple scatterings, we set it to unity in the following analysis to find an upper limit for the reflection.
Figure~\ref{fig:maxpol} shows the diluted PD $P_\mathrm{obs}$ and its maximum at each $d/r$.
Because the reflected flux and PD depend on the binary separation in opposite ways, the observed PD only goes up to a maximum of $\sim$2\% in the range $d/r \sim$1.2--1.6.
Compared to a Roche lobe this separation corresponds to small mass ratios of $q_\mathrm{m} \sim$0.005--0.1, and for $q_\mathrm{m}>1$ the polarization is less than one percent.
Accounting for the scattering albedo, the maximum at close separations is likely on the order of 0.1\%--0.7\%.
The PD increases at extremely small separations due to the geometry of the visible area approaching a plane, however, this is an unphysical scenario.

\begin{figure} 
\begin{center}
\includegraphics[width=0.8\linewidth]{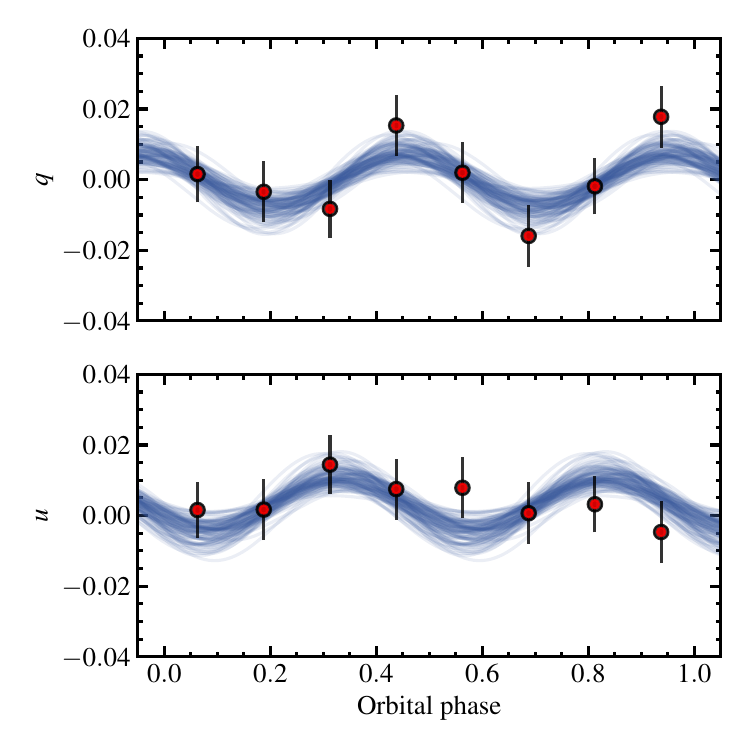}
\end{center}
\caption{Orbital variability of normalized Stokes parameters of \mbox{GS~1826$-$238} \citep{Rankin24} (red circles with 1$\sigma$ error bars). The solid blue lines show the large-separation reflection model given by Eqs.~\eqref{eq:normq} and \eqref{eq:normu} for 100 samples from the posterior distribution.   \label{fig:fit}} 
\end{figure}

\section{Applications} \label{sec:applications}

We fitted our model to existing IXPE data of the orbital polarization in \mbox{GS~1826$-$238}.
It features a weakly magnetized neutron star with near zero constant PD, making it ideal for the study of orbital polarization \citep{Capitano23}.
A previous study of the IXPE data by \citet{Rankin24} found that an optically thin reflection model describes the data better than assuming constant polarization.
Optical observations of the binary show a binary separation of $d/r \gtrsim 3$, and so the companion star can only cover a fraction of $\epsilon \lesssim 3\%$ of the sky \citep{Mescheryakov11} as seen from the X-ray source.
This sets an upper limit for the flux contributed by the stellar reflection alone, and the high flux fraction of $\lambda \epsilon = 2.7^{+1.0}_{-1.2} \%$ found by the \citet{Rankin24} fit implies that the reflection in \mbox{GS~1826$-$238} is likely very significant.
We performed the fit using Eqs.~\eqref{eq:normq} and \eqref{eq:normu} of the large-separation approximation due to its simplicity and the low accuracy of the data. 

The orbital solution for \mbox{GS~1826$-$238} is unknown, so we assumed a circular orbit and added a phase shift parameter $\omega$ to the orbital phase angle $\varphi$.
The model has four parameters: the inclination $i$, the reflected flux normalization $f_0$, the position angle of the orbital axis $\Omega$, and the phase shift $\omega$.
The observed normalized Stokes parameters are related to the theoretically computed in Sect.~\ref{sec:reflection} as :
\begin{align}
    q_\mathrm{obs} &= q \cos (2 \Omega) - u \sin (2 \Omega), \\
    u_\mathrm{obs} &= q \sin (2 \Omega) + u \cos (2 \Omega). 
\end{align}

\begin{figure} 
\begin{center}
\includegraphics[width=0.9\linewidth]{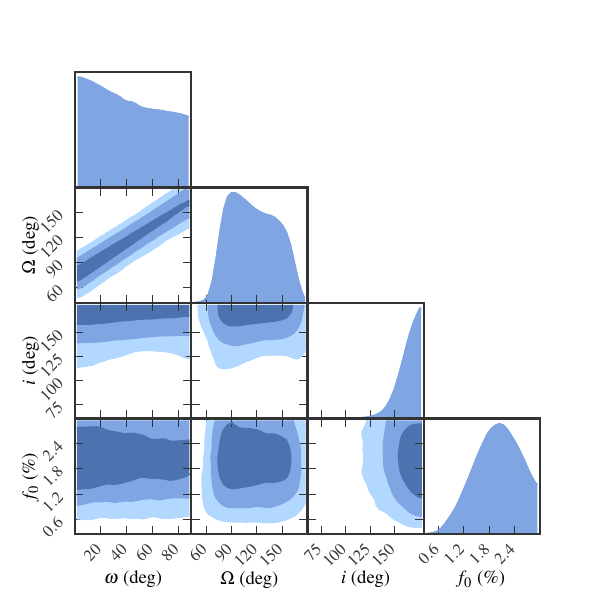}
\end{center}
\caption{Posterior distribution of the large-separation model parameters of the fit shown in Fig.~\ref{fig:fit}. The contours are 1, 2 and $3\sigma$. \label{fig:box}} 
\end{figure}

We employed Markov Chain Monte Carlo (MCMC) ensemble sampler implemented in \texttt{emcee} Python package \citep{emcee} to minimize the $\chi^2$ of the fit and to derive the posterior distributions for the model parameters. 
The best-fit model is presented in Fig.~\ref{fig:fit}, and its posteriors in Fig.~\ref{fig:box}.
The values of $i$, $\omega$, and $\Omega$ are consistent with the results from the optically thin model fit of \citet{Rankin24}, so the difference between the models may not be apparent within the accuracy of current data.
The optically thin fit is not sensitive for inclinations of $i \gtrsim 120\degr$, but our model fit shows a preference for inclinations close to $180\degr$.
Previously measured values for the inclination of \mbox{GS~1826$-$238} are $i = 62\fdg5 \pm 5\fdg5$ \citep{Mescheryakov11} and $i = 69^{+2}_{-3}$~deg \citep{Johnston20} (note the degeneracy between inclinations $i$ and $180\degr-i$ in those studies), so the reflection model does not seem to improve the constraints on the inclination.
The parameters $\omega$ and $\Omega$ are degenerate with one another and thus are difficult to constrain with no prior information.
Our fit of the scattering fraction $\lambda \epsilon \sim 5 \%$ is higher than the optically thin model by a factor of $\sim 2$, which is a consequence of the optically thin reflector predicting a much higher PD.
The amplitude of the observed variations cannot easily be explained by the stellar reflection model, especially as the accretion disk shadow further reduces the reflected flux.
The stellar reflection model produces sinusoidal variations of $q$ and $u$ only when the orbit is nearly edge-on, while the optically thin model is always sinusoidal.
The fit preferring inclinations near $180\degr$ could indicate that the reflecting medium is optically thin, although the evidence for this is inconclusive.

\section{Discussion} \label{sec:discussion}

Polarized X-ray reflection from the companion star is rather weak when diluted by direct emission from the compact object. 
Even though the single-scattering albedo is larger in the hard X-rays, the observed PD will remain less than 1\%.
The reflection should be most prominent in high-mass XRBs, since the observed polarization $P_\mathrm{obs}$ peaks at separations corresponding to the mass ratio $q_\mathrm{m}<1$. 
The dilution of the reflected light may be avoided if the direct emission from the compact source is blocked, while the companion star is visible. 
Because the opening angle of XRB accretion disks is about $10\degr$ \citep{Jong96}, an observer inclination over $\sim$80\degr\ can provide the necessary conditions for the eclipse of the central source. 
Alternatively, the direct emission can be blocked even at lower inclinations if the disk is warped. 

A famous example of an XRB with a warped disk is the X-ray pulsar \mbox{Her X-1}, which is viewed nearly edge-on \citep{Gerend76, Petterson75}.
It has a 35-day superorbital period with two 10-d long low states, during which the compact object is obscured by the accretion disk \citep{Scott00}. 
Similar disk obscuration is also seen in slightly lower inclination systems such as \mbox{LMC~X-4} and \mbox{SMC~X-1} \citep{Inoue19, Ogilvie01}.
However, all aforementioned targets are pulsars, whose direct emission is highly polarized \citep[e.g.,][]{Doroshenko2022,Doroshenko2023,Tsygankov2022,Tsygankov2023,Forsblom2023,Suleimanov2023,Mushtukov2023} and variable on a time scale much shorter than the orbital period, making detection of orbital variations related to the reflection from the companion an extremely difficult task. 
Some XRBs such as \mbox{SS~433}  and  \mbox{Cyg X-3} have thick equatorial obscurers that block the direct emission even at lower inclinations \citep{Fabrika04,Veledina23}, but the emission toward the star is also blocked. 
Orbital variations of X-ray polarization in \mbox{Cyg X-3} are then inconsistent with the reflection from the companion but rather consistent with reflection from some inhomogeneities in the stellar wind  \citep{Veledina23}.

Additionally, in dipping low-mass XRBs the accretion flow obscures the compact object from the observer near the eclipse \citep{Diaz06}, although the PD of the reflection would be low during this orbital phase and the accretion disk corona can still be visible.
IXPE observations of the dipping XRB \mbox{4U~1624$-$49} set an upper limit of 22\% for the PD during dips \citep{Saade24}.  
On the other hand, observations of a high-inclination weakly magnetized accreting neutron star \mbox{GX 13+1} revealed extremely complex variations of the polarization on timescales much below orbital period \citep{Bobrikova2024} also making detection of polarized reflection from the companion in such targets difficult. 

Our models do not account for the shadowing of the stellar surface by the accretion disk.
The shadowing reduces the reflected flux and overall reduces the amplitude of the orbital variations.
Consequently, our approximation is an upper limit for the variability.
The shadowing effect is most significant in XRBs with low-mass companions due to the small size of the star relative to the disk.
Depending on the exact geometry of the shadow it may not cover the area visible to the observer, especially if the disk is warped.
The orbital polarization will therefore be complex for XRBs with precessing warped disks.

Although the availability of observational data is limited, the amplitudes of the observed variations can be compared with the theoretical model.
Both \mbox{Cyg X-1} and \mbox{LMC~X-1} have high-mass companions, and thus their binary separation is on the order of $d/r \sim 2$.
The nondetection of orbital polarization variations in \mbox{Cyg X-1} \citep{Krawczynki21} means that any variability must be smaller than the statistical noise, which is in line with our predictions.
Although \mbox{LMC X-1} was not observed over many orbital periods, the data are consistent with PD variations of a few percent \citep{Podgorny23}.
Assuming this detection is reliable, our stellar reflection model cannot realistically produce PDs this high.
The X-ray light curve of the \mbox{LMC X-1} is modulated by about 7\%  which is consistent with electron scattering in the stellar wind, so the polarization is likely also dominated by wind scattering \citep{Orosz09}.
Our fit of the \mbox{GS~1826$-$238} data similarly shows a need for an unexpectedly high reflected flux, so either the direct emission is partially obscured, the emission from the central source is anisotropic, or the polarization is dominated by some other component.
This component could be the disk, the bulge where the accretion stream hits the disk, or the wind. 
Understanding the nature of the variability of the polarization in this source requires more data. 

\section{Conclusions}

We developed analytical models for the polarized X-ray reflection from the stellar companion in XRBs and performed fits to the existing data on the low-mass X-ray binary \mbox{GS~1826$-$238}.
The quality of the data is not sufficient to constrain the orbital parameters, but we find that the observed amplitude of the variations of the Stokes parameters is surprisingly large.
If diluted by the direct emission, the reflection from the companion's stellar surface cannot typically produce the observed PD of more than 1\% (unless the source is anisotropic), making the detection difficult under most circumstances.
The amplitude of the variability in both \mbox{GS~1826$-$238} and \mbox{LMC X-1} appears greater than what stellar reflection could produce.
The polarized reflection is expected to be more noticeable in XRBs where the direct emission is obscured, although many such targets are pulsars with highly variable compact object emission.
Higher-quality data are required to determine the origin of the orbital polarization.

\begin{acknowledgements}
This research has been supported by the Academy of Finland grants 333112 (JP),  355672 (VA), and by Finnish Cultural Foundation (VK). 
\end{acknowledgements}


\bibliographystyle{aa}
\bibliography{star}

\begin{thebibliography}{47}
\expandafter\ifx\csname natexlab\endcsname\relax\def\natexlab#1{#1}\fi

\bibitem[{{Abdul Qadir} {et~al.}(2023){Abdul Qadir}, {Berdyugin}, {Piirola},
  {Sakanoi}, \& {Kagitani}}]{AbdulQadir23}
{Abdul Qadir}, Y., {Berdyugin}, A.~V., {Piirola}, V., {Sakanoi}, T., \&
  {Kagitani}, M. 2023, \aap, 677, A75

\bibitem[{{Basko} {et~al.}(1974){Basko}, {Sunyaev}, \& {Titarchuk}}]{BS74}
{Basko}, M.~M., {Sunyaev}, R.~A., \& {Titarchuk}, L.~G. 1974, \aap, 31, 249

\bibitem[{{Berdyugin} {et~al.}(2016){Berdyugin}, {Piirola}, {Sadegi},
  {Tsygankov}, {Sakanoi}, {Kagitani}, {Yoneda}, {Okano}, \&
  {Poutanen}}]{Berdyugin16}
{Berdyugin}, A., {Piirola}, V., {Sadegi}, S., {et~al.} 2016, \aap, 591, A92

\bibitem[{{Berdyugin} {et~al.}(2018){Berdyugin}, {Piirola}, {Sakanoi},
  {Kagitani}, \& {Yoneda}}]{Berdyugin18}
{Berdyugin}, A., {Piirola}, V., {Sakanoi}, T., {Kagitani}, M., \& {Yoneda}, M.
  2018, \aap, 611, A69

\bibitem[{{Berdyugina} {et~al.}(2011){Berdyugina}, {Berdyugin}, {Fluri}, \&
  {Piirola}}]{Berdyugina11}
{Berdyugina}, S.~V., {Berdyugin}, A.~V., {Fluri}, D.~M., \& {Piirola}, V. 2011,
  \apjl, 728, L6

\bibitem[{{Blondin}(1994)}]{Blondin94}
{Blondin}, J.~M. 1994, \apj, 435, 756

\bibitem[{{Bobrikova} {et~al.}(2024){Bobrikova}, {Forsblom}, {Di Marco}, {La
  Monaca}, {Poutanen}, {Ng}, {Ravi}, {Loktev}, {Kajava}, {Ursini}, {Veledina},
  {Rogantini}, {Salmi}, {Bianchi}, {Capitanio}, {Done}, {Fabiani}, {Gnarini},
  {Heyl}, {Kaaret}, {Matt}, {Muleri}, {Nitindala}, {Rankin}, {Weisskopf},
  {Agudo}, {Antonelli}, {Bachetti}, {Baldini}, {Baumgartner}, {Bellazzini},
  {Bongiorno}, {Bonino}, {Brez}, {Bucciantini}, {Castellano}, {Cavazzuti},
  {Chen}, {Ciprini}, {Costa}, {De Rosa}, {Del Monte}, {Di Gesu}, {Di Lalla},
  {Donnarumma}, {Doroshenko}, {Dovciak}, {Ehlert}, {Enoto}, {Evangelista},
  {Ferrazzoli}, {Garcia}, {Gunji}, {Hayashida}, {Iwakiri}, {Jorstad}, {Karas},
  {Kislat}, {Kitaguchi}, {Kolodziejczak}, {Krawczynski}, {Latronico},
  {Liodakis}, {Maldera}, {Manfreda}, {Marin}, {Marinucci}, {Marscher},
  {Marshall}, {Massaro}, {Mitsuishi}, {Mizuno}, {Negro}, {Ng}, {O'Dell},
  {Omodei}, {Oppedisano}, {Papitto}, {Pavlov}, {Peirson}, {Perri},
  {Pesce-Rollins}, {Petrucci}, {Pilia}, {Possenti}, {Puccetti}, {Ramsey},
  {Ratheesh}, {Roberts}, {Romani}, {Sgro}, {Slane}, {Soffitta}, {Spandre},
  {Swartz}, {Tamagawa}, {Tavecchio}, {Taverna}, {Tawara}, {Tennant}, {Thomas},
  {Tombesi}, {Trois}, {Tsygankov}, {Turolla}, {Vink}, {Wu}, {Xie}, \&
  {Zane}}]{Bobrikova2024}
{Bobrikova}, A., {Forsblom}, S.~V., {Di Marco}, A., {et~al.} 2024, \aap,
  submitted, arXiv:2401.13058

\bibitem[{{Brown} {et~al.}(1978){Brown}, {McLean}, \& {Emslie}}]{Brown78}
{Brown}, J.~C., {McLean}, I.~S., \& {Emslie}, A.~G. 1978, \aap, 68, 415

\bibitem[{{Capitanio} {et~al.}(2023){Capitanio}, {Fabiani}, {Gnarini},
  {Ursini}, {Ferrigno}, {Matt}, {Poutanen}, {Cocchi}, {Mikusincova},
  {Farinelli}, {Bianchi}, {Kajava}, {Muleri}, {Sanchez-Fernandez}, {Soffitta},
  {Wu}, {Agudo}, {Antonelli}, {Bachetti}, {Baldini}, {Baumgartner},
  {Bellazzini}, {Bongiorno}, {Bonino}, {Brez}, {Bucciantini}, {Castellano},
  {Cavazzuti}, {Ciprini}, {Costa}, {De Rosa}, {Del Monte}, {Di Gesu}, {Di
  Lalla}, {Di Marco}, {Donnarumma}, {Doroshenko}, {Dov{\v{c}}iak}, {Ehlert},
  {Enoto}, {Evangelista}, {Ferrazzoli}, {Garcia}, {Gunji}, {Hayashida}, {Heyl},
  {Iwakiri}, {Jorstad}, {Karas}, {Kitaguchi}, {Kolodziejczak}, {Krawczynski},
  {La Monaca}, {Latronico}, {Liodakis}, {Maldera}, {Manfreda}, {Marin},
  {Marinucci}, {Marscher}, {Marshall}, {Mitsuishi}, {Mizuno}, {Ng}, {O'Dell},
  {Omodei}, {Oppedisano}, {Papitto}, {Pavlov}, {Peirson}, {Perri},
  {Pesce-Rollins}, {Petrucci}, {Pilia}, {Possenti}, {Puccetti}, {Ramsey},
  {Rankin}, {Ratheesh}, {Romani}, {Sgr{\`o}}, {Slane}, {Spandre}, {Tamagawa},
  {Tavecchio}, {Taverna}, {Tawara}, {Tennant}, {Thomas}, {Tombesi}, {Trois},
  {Tsygankov}, {Turolla}, {Vink}, {Weisskopf}, {Xie}, \& {Zane}}]{Capitano23}
{Capitanio}, F., {Fabiani}, S., {Gnarini}, A., {et~al.} 2023, \apj, 943, 129

\bibitem[{{Chandrasekhar}(1960)}]{Cha60}
{Chandrasekhar}, S. 1960, {Radiative Transfer} (New York: Dover)

\bibitem[{{de Jong} {et~al.}(1996){de Jong}, {van Paradijs}, \&
  {Augusteijn}}]{Jong96}
{de Jong}, J.~A., {van Paradijs}, J., \& {Augusteijn}, T. 1996, \aap, 314, 484

\bibitem[{{D{\'\i}az Trigo} {et~al.}(2006){D{\'\i}az Trigo}, {Parmar},
  {Boirin}, {M{\'e}ndez}, \& {Kaastra}}]{Diaz06}
{D{\'\i}az Trigo}, M., {Parmar}, A.~N., {Boirin}, L., {M{\'e}ndez}, M., \&
  {Kaastra}, J.~S. 2006, \aap, 445, 179

\bibitem[{{Dolan} \& {Tapia}(1989{\natexlab{a}})}]{Dolan89A0620}
{Dolan}, J.~F. \& {Tapia}, S. 1989{\natexlab{a}}, \pasp, 101, 1135

\bibitem[{{Dolan} \& {Tapia}(1989{\natexlab{b}})}]{Dolan89CygX1}
{Dolan}, J.~F. \& {Tapia}, S. 1989{\natexlab{b}}, \apj, 344, 830

\bibitem[{{Doroshenko} {et~al.}(2023){Doroshenko}, {Poutanen}, {Heyl},
  {Tsygankov}, {Caiazzo}, {Turolla}, {Veledina}, {Weisskopf}, {Forsblom},
  {Gonz{\'a}lez-Caniulef}, {Loktev}, {Malacaria}, {Mushtukov}, {Suleimanov},
  {Lutovinov}, {Mereminskiy}, {Molkov}, {Salganik}, {Santangelo}, {Berdyugin},
  {Kravtsov}, {Nitindala}, {Agudo}, {Antonelli}, {Bachetti}, {Baldini},
  {Baumgartner}, {Bellazzini}, {Bianchi}, {Bongiorno}, {Bonino}, {Brez},
  {Bucciantini}, {Capitanio}, {Castellano}, {Cavazzuti}, {Chen}, {Ciprini},
  {Costa}, {De Rosa}, {Del Monte}, {Di Gesu}, {Di Lalla}, {Di Marco},
  {Donnarumma}, {Dov{\v{c}}iak}, {Ehlert}, {Enoto}, {Evangelista}, {Fabiani},
  {Ferrazzoli}, {Garc{\'\i}a}, {Gunji}, {Hayashida}, {Iwakiri}, {Jorstad},
  {Kaaret}, {Karas}, {Kislat}, {Kitaguchi}, {Kolodziejczak}, {Krawczynski}, {La
  Monaca}, {Latronico}, {Liodakis}, {Maldera}, {Manfreda}, {Marin},
  {Marinucci}, {Marscher}, {Marshall}, {Massaro}, {Matt}, {Mitsuishi},
  {Mizuno}, {Muleri}, {Negro}, {Ng}, {O'Dell}, {Omodei}, {Oppedisano},
  {Papitto}, {Pavlov}, {Peirson}, {Perri}, {Pesce-Rollins}, {Petrucci},
  {Pilia}, {Possenti}, {Puccetti}, {Ramsey}, {Rankin}, {Ratheesh}, {Roberts},
  {Romani}, {Sgr{\`o}}, {Slane}, {Soffitta}, {Spandre}, {Swartz}, {Tamagawa},
  {Tavecchio}, {Taverna}, {Tawara}, {Tennant}, {Thomas}, {Tombesi}, {Trois},
  {Vink}, {Wu}, {Xie}, \& {Zane}}]{Doroshenko2023}
{Doroshenko}, V., {Poutanen}, J., {Heyl}, J., {et~al.} 2023, \aap, 677, A57

\bibitem[{{Doroshenko} {et~al.}(2022){Doroshenko}, {Poutanen}, {Tsygankov},
  {Suleimanov}, {Bachetti}, {Caiazzo}, {Costa}, {Di Marco}, {Heyl}, {La
  Monaca}, {Muleri}, {Mushtukov}, {Pavlov}, {Ramsey}, {Rankin}, {Santangelo},
  {Soffitta}, {Staubert}, {Weisskopf}, {Zane}, {Agudo}, {Antonelli}, {Baldini},
  {Baumgartner}, {Bellazzini}, {Bianchi}, {Bongiorno}, {Bonino}, {Brez},
  {Bucciantini}, {Capitanio}, {Castellano}, {Cavazzuti}, {Ciprini}, {De Rosa},
  {Del Monte}, {Di Gesu}, {Di Lalla}, {Donnarumma}, {Dov{\v{c}}iak}, {Ehlert},
  {Enoto}, {Evangelista}, {Fabiani}, {Ferrazzoli}, {Garcia}, {Gunji},
  {Hayashida}, {Iwakiri}, {Jorstad}, {Karas}, {Kitaguchi}, {Kolodziejczak},
  {Krawczynski}, {Latronico}, {Liodakis}, {Maldera}, {Manfreda}, {Marin},
  {Marinucci}, {Marscher}, {Marshall}, {Matt}, {Mitsuishi}, {Mizuno}, {Ng},
  {O'Dell}, {Omodei}, {Oppedisano}, {Papitto}, {Peirson}, {Perri},
  {Pesce-Rollins}, {Pilia}, {Possenti}, {Puccetti}, {Ratheesh}, {Romani},
  {Sgr{\`o}}, {Slane}, {Spandre}, {Sunyaev}, {Tamagawa}, {Tavecchio},
  {Taverna}, {Tawara}, {Tennant}, {Thomas}, {Tombesi}, {Trois}, {Turolla},
  {Vink}, {Wu}, \& {Xie}}]{Doroshenko2022}
{Doroshenko}, V., {Poutanen}, J., {Tsygankov}, S.~S., {et~al.} 2022, Nature
  Astronomy, 6, 1433

\bibitem[{{Eggleton}(1983)}]{Eggleton83}
{Eggleton}, P.~P. 1983, \apj, 268, 368

\bibitem[{{Fabrika}(2004)}]{Fabrika04}
{Fabrika}, S. 2004, Astrophysics and Space Physics Reviews, 12, 1

\bibitem[{{Foreman-Mackey} {et~al.}(2013){Foreman-Mackey}, {Hogg}, {Lang}, \&
  {Goodman}}]{emcee}
{Foreman-Mackey}, D., {Hogg}, D.~W., {Lang}, D., \& {Goodman}, J. 2013, PASP,
  125, 306

\bibitem[{{Forsblom} {et~al.}(2023){Forsblom}, {Poutanen}, {Tsygankov},
  {Bachetti}, {Di Marco}, {Doroshenko}, {Heyl}, {La Monaca}, {Malacaria},
  {Marshall}, {Muleri}, {Mushtukov}, {Pilia}, {Rogantini}, {Suleimanov},
  {Taverna}, {Xie}, {Agudo}, {Antonelli}, {Baldini}, {Baumgartner},
  {Bellazzini}, {Bianchi}, {Bongiorno}, {Bonino}, {Brez}, {Bucciantini},
  {Capitanio}, {Castellano}, {Cavazzuti}, {Chen}, {Ciprini}, {Costa}, {De
  Rosa}, {Del Monte}, {Di Gesu}, {Di Lalla}, {Donnarumma}, {Dov{\v{c}}iak},
  {Ehlert}, {Enoto}, {Evangelista}, {Fabiani}, {Ferrazzoli}, {Garcia}, {Gunji},
  {Hayashida}, {Iwakiri}, {Jorstad}, {Kaaret}, {Karas}, {Kitaguchi},
  {Kolodziejczak}, {Krawczynski}, {Latronico}, {Liodakis}, {Maldera},
  {Manfreda}, {Marin}, {Marinucci}, {Marscher}, {Matt}, {Mitsuishi}, {Mizuno},
  {Negro}, {Ng}, {O'Dell}, {Omodei}, {Oppedisano}, {Papitto}, {Pavlov},
  {Peirson}, {Perri}, {Pesce-Rollins}, {Petrucci}, {Possenti}, {Puccetti},
  {Ramsey}, {Rankin}, {Ratheesh}, {Roberts}, {Romani}, {Sgr{\`o}}, {Slane},
  {Soffitta}, {Spandre}, {Sunyaev}, {Swartz}, {Tamagawa}, {Tavecchio},
  {Tawara}, {Tennant}, {Thomas}, {Tombesi}, {Trois}, {Turolla}, {Vink},
  {Weisskopf}, {Wu}, {Zane}, \& {IXPE Collaboration}}]{Forsblom2023}
{Forsblom}, S.~V., {Poutanen}, J., {Tsygankov}, S.~S., {et~al.} 2023, \apjl,
  947, L20

\bibitem[{{Frank} {et~al.}(2002){Frank}, {King}, \& {Raine}}]{Frank02}
{Frank}, J., {King}, A., \& {Raine}, D.~J. 2002, {Accretion Power in
  Astrophysics} (Cambridge: Cambridge University Press)

\bibitem[{{Gerend} \& {Boynton}(1976)}]{Gerend76}
{Gerend}, D. \& {Boynton}, P.~E. 1976, \apj, 209, 562

\bibitem[{{Gnedin} \& {Sunyaev}(1974)}]{GS74}
{Gnedin}, I.~N. \& {Sunyaev}, R.~A. 1974, \aap, 36, 379

\bibitem[{{Inoue}(2019)}]{Inoue19}
{Inoue}, H. 2019, \pasj, 71, 36

\bibitem[{{Johnston} {et~al.}(2020){Johnston}, {Heger}, \&
  {Galloway}}]{Johnston20}
{Johnston}, Z., {Heger}, A., \& {Galloway}, D.~K. 2020, \mnras, 494, 4576

\bibitem[{{Kemp} {et~al.}(1978){Kemp}, {Barbour}, {Herman}, \& {Rudy}}]{Kemp78}
{Kemp}, J.~C., {Barbour}, M.~S., {Herman}, L.~C., \& {Rudy}, R.~J. 1978, \apjl,
  220, L123

\bibitem[{{Kravtsov} {et~al.}(2020){Kravtsov}, {Berdyugin}, {Piirola},
  {Kosenkov}, {Tsygankov}, {Chernyakova}, {Malyshev}, {Sakanoi}, {Kagitani},
  {Berdyugina}, \& {Poutanen}}]{Kravtsov20}
{Kravtsov}, V., {Berdyugin}, A.~V., {Piirola}, V., {et~al.} 2020, \aap, 643,
  A170

\bibitem[{{Kravtsov} {et~al.}(2023){Kravtsov}, {Veledina}, {Berdyugin},
  {Zdziarski}, {Henson}, {Piirola}, {Sakanoi}, {Kagitani}, {Berdyugina}, \&
  {Poutanen}}]{Kravtsov23}
{Kravtsov}, V., {Veledina}, A., {Berdyugin}, A.~V., {et~al.} 2023, \aap, 678,
  A58

\bibitem[{{Krawczynski} {et~al.}(2022){Krawczynski}, {Muleri}, {Dov{\v{c}}iak},
  {Veledina}, {Rodriguez Cavero}, {Svoboda}, {Ingram}, {Matt}, {Garcia},
  {Loktev}, {Negro}, {Poutanen}, {Kitaguchi}, {Podgorn{\'y}}, {Rankin},
  {Zhang}, {Berdyugin}, {Berdyugina}, {Bianchi}, {Blinov}, {Capitanio}, {Di
  Lalla}, {Draghis}, {Fabiani}, {Kagitani}, {Kravtsov}, {Kiehlmann},
  {Latronico}, {Lutovinov}, {Mandarakas}, {Marin}, {Marinucci}, {Miller},
  {Mizuno}, {Molkov}, {Omodei}, {Petrucci}, {Ratheesh}, {Sakanoi}, {Semena},
  {Skalidis}, {Soffitta}, {Tennant}, {Thalhammer}, {Tombesi}, {Weisskopf},
  {Wilms}, {Zhang}, {Agudo}, {Antonelli}, {Bachetti}, {Baldini}, {Baumgartner},
  {Bellazzini}, {Bongiorno}, {Bonino}, {Brez}, {Bucciantini}, {Castellano},
  {Cavazzuti}, {Ciprini}, {Costa}, {De Rosa}, {Del Monte}, {Di Gesu}, {Di
  Marco}, {Donnarumma}, {Doroshenko}, {Ehlert}, {Enoto}, {Evangelista},
  {Ferrazzoli}, {Gunji}, {Hayashida}, {Heyl}, {Iwakiri}, {Jorstad}, {Karas},
  {Kolodziejczak}, {La Monaca}, {Liodakis}, {Maldera}, {Manfreda}, {Marscher},
  {Marshall}, {Mitsuishi}, {Ng}, {O{\textquoteright}Dell}, {Oppedisano},
  {Papitto}, {Pavlov}, {Peirson}, {Perri}, {Pesce-Rollins}, {Pilia},
  {Possenti}, {Puccetti}, {Ramsey}, {Romani}, {Sgr{\`o}}, {Slane}, {Spandre},
  {Tamagawa}, {Tavecchio}, {Taverna}, {Tawara}, {Thomas}, {Trois}, {Tsygankov},
  {Turolla}, {Vink}, {Wu}, {Xie}, \& {Zane}}]{Krawczynki21}
{Krawczynski}, H., {Muleri}, F., {Dov{\v{c}}iak}, M., {et~al.} 2022, Science,
  378, 650

\bibitem[{{Leahy} \& {Leahy}(2015)}]{Leahy2015}
{Leahy}, D.~A. \& {Leahy}, J.~C. 2015, Computational Astrophysics and
  Cosmology, 2, 4

\bibitem[{{Madhusudhan} \& {Burrows}(2012)}]{MB12}
{Madhusudhan}, N. \& {Burrows}, A. 2012, \apj, 747, 25

\bibitem[{{Mescheryakov} {et~al.}(2011){Mescheryakov}, {Revnivtsev}, \&
  {Filippova}}]{Mescheryakov11}
{Mescheryakov}, A.~V., {Revnivtsev}, M.~G., \& {Filippova}, E.~V. 2011,
  Astronomy Letters, 37, 826

\bibitem[{{Mushtukov} {et~al.}(2023){Mushtukov}, {Tsygankov}, {Poutanen},
  {Doroshenko}, {Salganik}, {Costa}, {Marco}, {Heyl}, {Monaca}, {Lutovinov},
  {Mereminsky}, {Papitto}, {Semena}, {Shtykovsky}, {Suleimanov}, {Forsblom},
  {Gonz{\'a}lez-Caniulef}, {Malacaria}, {Sunyaev}, {Agudo}, {Antonelli},
  {Bachetti}, {Baldini}, {Baumgartner}, {Bellazzini}, {Bianchi}, {Bongiorno},
  {Bonino}, {Brez}, {Bucciantini}, {Capitanio}, {Castellano}, {Cavazzuti},
  {Chen}, {Ciprini}, {De Rosa}, {Del Monte}, {Gesu}, {Lalla}, {Donnarumma},
  {Dov{\v{c}}iak}, {Ehlert}, {Enoto}, {Evangelista}, {Fabiani}, {Ferrazzoli},
  {Garcia}, {Gunji}, {Hayashida}, {Iwakiri}, {Jorstad}, {Kaaret}, {Karas},
  {Kislat}, {Kitaguchi}, {Kolodziejczak}, {Krawczynski}, {Latronico},
  {Liodakis}, {Maldera}, {Manfreda}, {Marin}, {Marscher}, {Marshall},
  {Massaro}, {Matt}, {Mitsuishi}, {Mizuno}, {Muleri}, {Negro}, {Ng}, {O'Dell},
  {Omodei}, {Oppedisano}, {Pavlov}, {Peirson}, {Perri}, {Pesce-Rollins},
  {Petrucci}, {Pilia}, {Possenti}, {Puccetti}, {Ramsey}, {Rankin}, {Ratheesh},
  {Roberts}, {Romani}, {Sgr{\`o}}, {Slane}, {Soffitta}, {Spandre}, {Swartz},
  {Tamagawa}, {Tavecchio}, {Taverna}, {Tawara}, {Tennant}, {Thomas}, {Tombesi},
  {Trois}, {Turolla}, {Vink}, {Weisskopf}, {Wu}, {Xie}, \&
  {Zane}}]{Mushtukov2023}
{Mushtukov}, A.~A., {Tsygankov}, S.~S., {Poutanen}, J., {et~al.} 2023, \mnras,
  524, 2004

\bibitem[{{Ogilvie} \& {Dubus}(2001)}]{Ogilvie01}
{Ogilvie}, G.~I. \& {Dubus}, G. 2001, \mnras, 320, 485

\bibitem[{{Orosz} {et~al.}(2009){Orosz}, {Steeghs}, {McClintock}, {Torres},
  {Bochkov}, {Gou}, {Narayan}, {Blaschak}, {Levine}, {Remillard}, {Bailyn},
  {Dwyer}, \& {Buxton}}]{Orosz09}
{Orosz}, J.~A., {Steeghs}, D., {McClintock}, J.~E., {et~al.} 2009, \apj, 697,
  573

\bibitem[{{Petterson}(1975)}]{Petterson75}
{Petterson}, J.~A. 1975, \apjl, 201, L61

\bibitem[{{Podgorn{\'y}} {et~al.}(2023){Podgorn{\'y}}, {Marra}, {Muleri},
  {Rodriguez Cavero}, {Ratheesh}, {Dov{\v{c}}iak}, {Miku{\v{s}}incov{\'a}},
  {Brigitte}, {Steiner}, {Veledina}, {Bianchi}, {Krawczynski}, {Svoboda},
  {Kaaret}, {Matt}, {Garc{\'\i}a}, {Petrucci}, {Lutovinov}, {Semena}, {Di
  Marco}, {Negro}, {Weisskopf}, {Ingram}, {Poutanen}, {Beheshtipour}, {Chun},
  {Hu}, {Mizuno}, {Sixuan}, {Tombesi}, {Zane}, {Agudo}, {Antonelli},
  {Bachetti}, {Baldini}, {Baumgartner}, {Bellazzini}, {Bongiorno}, {Bonino},
  {Brez}, {Bucciantini}, {Capitanio}, {Castellano}, {Cavazzuti}, {Chen},
  {Ciprini}, {Costa}, {De Rosa}, {Del Monte}, {Di Gesu}, {Di Lalla},
  {Donnarumma}, {Doroshenko}, {Ehlert}, {Enoto}, {Evangelista}, {Fabiani},
  {Ferrazzoli}, {Gunji}, {Hayashida}, {Heyl}, {Iwakiri}, {Jorstad}, {Karas},
  {Kislat}, {Kitaguchi}, {Kolodziejczak}, {La Monaca}, {Latronico}, {Liodakis},
  {Maldera}, {Manfreda}, {Marin}, {Marinucci}, {Marscher}, {Marshall},
  {Massaro}, {Mitsuishi}, {Ng}, {O'Dell}, {Omodei}, {Oppedisano}, {Papitto},
  {Pavlov}, {Peirson}, {Perri}, {Pesce-Rollins}, {Pilia}, {Possenti},
  {Puccetti}, {Ramsey}, {Rankin}, {Roberts}, {Romani}, {Sgr{\`o}}, {Slane},
  {Soffitta}, {Spandre}, {Swartz}, {Tamagawa}, {Tavecchio}, {Taverna},
  {Tawara}, {Tennant}, {Thomas}, {Trois}, {Tsygankov}, {Turolla}, {Vink}, {Wu},
  \& {Xie}}]{Podgorny23}
{Podgorn{\'y}}, J., {Marra}, L., {Muleri}, F., {et~al.} 2023, \mnras, 526, 5964

\bibitem[{{Rankin} {et~al.}(2024){Rankin}, {Kravtsov}, {Muleri}, {Poutanen},
  {Marin}, {Capitanio}, {Matt}, {Costa}, {Di Marco}, {Fabiani}, {La Monaca},
  {Marra}, \& {Soffitta}}]{Rankin24}
{Rankin}, J., {Kravtsov}, V., {Muleri}, F., {et~al.} 2024, \apj, 962, 34

\bibitem[{{Russell}(1916)}]{Russell1916}
{Russell}, H.~N. 1916, \apj, 43, 173

\bibitem[{{Saade} {et~al.}(2024){Saade}, {Kaaret}, {Gnarini}, {Poutanen},
  {Ursini}, {Bianchi}, {Bobrikova}, {La Monaca}, {Di Marco}, {Capitanio},
  {Veledina}, {Agudo}, {Antonelli}, {Bachetti}, {Baldini}, {Baumgartner},
  {Bellazzini}, {Bongiorno}, {Bonino}, {Brez}, {Bucciantini}, {Castellano},
  {Cavazzuti}, {Chen}, {Ciprini}, {Costa}, {De Rosa}, {Del Monte}, {Di
  Ges{\`u}}, {Di Lalla}, {Donnarumma}, {Doroshenko}, {Dovciak}, {Ehlert},
  {Emote}, {Evangelista}, {Fabiani}, {Ferrazzoli}, {Garcia}, {Gunji},
  {Hayashida}, {Heyl}, {Iwakiri}, {Jorstad}, {Karas}, {Kislat}, {Kitaguchi},
  {Kolodziejczak}, {Krawczynski}, {Latronico}, {Liodakis}, {Maldera},
  {Manfreda}, {Marin}, {Marinucci}, {Marscher}, {Marshall}, {Massaro}, {Matt},
  {Mitsuishi}, {Mizudo}, {Muleri}, {Negro}, {Ng}, {O'Dell}, {Omodei},
  {Oppedisano}, {Papitto}, {Pavlov}, {Peirson}, {Perri}, {Pesce-Rollins},
  {Petrucci}, {Pilia}, {Possenti}, {Puccetti}, {Ramsey}, {Rankin}, {Ratheesh},
  {Roberts}, {Romani}, {Sgro}, {Slane}, {Soffitta}, {Spandre}, {Zwartz},
  {Tamagawa}, {Tavecchio}, {Taverna}, {Tawara}, {Tenant}, {Thomas}, {Tombesi},
  {Trois}, {Tsygankov}, {Turolla}, {Vink}, {Weisskopf}, {Wu}, {Xie}, \&
  {Zane}}]{Saade24}
{Saade}, M.~L., {Kaaret}, P., {Gnarini}, A., {et~al.} 2024, \apj, 963, 133

\bibitem[{{Scott} {et~al.}(2000){Scott}, {Leahy}, \& {Wilson}}]{Scott00}
{Scott}, D.~M., {Leahy}, D.~A., \& {Wilson}, R.~B. 2000, \apj, 539, 392

\bibitem[{{Sobolev}(1975)}]{Sobolev75}
{Sobolev}, V.~V. 1975, {Light scattering in planetary atmospheres} (Oxford:
  Pergamon Press)

\bibitem[{{Suleimanov} {et~al.}(2023){Suleimanov}, {Forsblom}, {Tsygankov},
  {Poutanen}, {Doroshenko}, {Doroshenko}, {Capitanio}, {Di Marco},
  {Gonz{\'a}lez-Caniulef}, {Heyl}, {La Monaca}, {Lutovinov}, {Molkov},
  {Malacaria}, {Mushtukov}, {Shtykovsky}, {Agudo}, {Antonelli}, {Bachetti},
  {Baldini}, {Baumgartner}, {Bellazzini}, {Bianchi}, {Bongiorno}, {Bonino},
  {Brez}, {Bucciantini}, {Castellano}, {Cavazzuti}, {Chen}, {Ciprini}, {Costa},
  {De Rosa}, {Del Monte}, {Di Gesu}, {Di Lalla}, {Donnarumma}, {Dov{\v{c}}iak},
  {Ehlert}, {Enoto}, {Evangelista}, {Fabiani}, {Ferrazzoli}, {Garcia}, {Gunji},
  {Hayashida}, {Iwakiri}, {Jorstad}, {Kaaret}, {Karas}, {Kislat}, {Kitaguchi},
  {Kolodziejczak}, {Krawczynski}, {Latronico}, {Liodakis}, {Maldera},
  {Manfreda}, {Marin}, {Marinucci}, {Marscher}, {Marshall}, {Massaro}, {Matt},
  {Mitsuishi}, {Mizuno}, {Muleri}, {Negro}, {Ng}, {O'Dell}, {Omodei},
  {Oppedisano}, {Papitto}, {Pavlov}, {Peirson}, {Perri}, {Pesce-Rollins},
  {Petrucci}, {Pilia}, {Possenti}, {Puccetti}, {Ramsey}, {Rankin}, {Ratheesh},
  {Roberts}, {Romani}, {Sgr{\`o}}, {Slane}, {Soffitta}, {Spandre}, {Swartz},
  {Tamagawa}, {Tavecchio}, {Taverna}, {Tawara}, {Tennant}, {Thomas}, {Tombesi},
  {Trois}, {Turolla}, {Vink}, {Weisskopf}, {Wu}, {Xie}, \&
  {Zane}}]{Suleimanov2023}
{Suleimanov}, V.~F., {Forsblom}, S.~V., {Tsygankov}, S.~S., {et~al.} 2023,
  \aap, 678, A119

\bibitem[{{Tsygankov} {et~al.}(2023){Tsygankov}, {Doroshenko}, {Mushtukov},
  {Poutanen}, {Di Marco}, {Heyl}, {La Monaca}, {Forsblom}, {Malacaria},
  {Marshall}, {Suleimanov}, {Svoboda}, {Taverna}, {Ursini}, {Agudo},
  {Antonelli}, {Bachetti}, {Baldini}, {Baumgartner}, {Bellazzini}, {Bianchi},
  {Bongiorno}, {Bonino}, {Brez}, {Bucciantini}, {Capitanio}, {Castellano},
  {Cavazzuti}, {Chen}, {Ciprini}, {Costa}, {De Rosa}, {Del Monte}, {Di Gesu},
  {Di Lalla}, {Donnarumma}, {Dov{\v{c}}iak}, {Ehlert}, {Enoto}, {Evangelista},
  {Fabiani}, {Ferrazzoli}, {Garcia}, {Gunji}, {Hayashida}, {Iwakiri},
  {Jorstad}, {Kaaret}, {Karas}, {Kislat}, {Kitaguchi}, {Kolodziejczak},
  {Krawczynski}, {Latronico}, {Liodakis}, {Maldera}, {Manfreda}, {Marin},
  {Marinucci}, {Marscher}, {Massaro}, {Matt}, {Mitsuishi}, {Mizuno}, {Muleri},
  {Negro}, {Ng}, {O'Dell}, {Omodei}, {Oppedisano}, {Papitto}, {Pavlov},
  {Peirson}, {Perri}, {Pesce-Rollins}, {Petrucci}, {Pilia}, {Possenti},
  {Puccetti}, {Ramsey}, {Rankin}, {Ratheesh}, {Roberts}, {Romani}, {Sgr{\`o}},
  {Slane}, {Soffitta}, {Spandre}, {Swartz}, {Tamagawa}, {Tavecchio}, {Tawara},
  {Tennant}, {Thomas}, {Tombesi}, {Trois}, {Turolla}, {Vink}, {Weisskopf},
  {Wu}, {Xie}, \& {Zane}}]{Tsygankov2023}
{Tsygankov}, S.~S., {Doroshenko}, V., {Mushtukov}, A.~A., {et~al.} 2023, \aap,
  675, A48

\bibitem[{{Tsygankov} {et~al.}(2022){Tsygankov}, {Doroshenko}, {Poutanen},
  {Heyl}, {Mushtukov}, {Caiazzo}, {Di Marco}, {Forsblom},
  {Gonz{\'a}lez-Caniulef}, {Klawin}, {La Monaca}, {Malacaria}, {Marshall},
  {Muleri}, {Ng}, {Suleimanov}, {Sunyaev}, {Turolla}, {Agudo}, {Antonelli},
  {Bachetti}, {Baldini}, {Baumgartner}, {Bellazzini}, {Bianchi}, {Bongiorno},
  {Bonino}, {Brez}, {Bucciantini}, {Capitanio}, {Castellano}, {Cavazzuti},
  {Ciprini}, {Costa}, {De Rosa}, {Del Monte}, {Di Gesu}, {Di Lalla},
  {Donnarumma}, {Dov{\v{c}}iak}, {Ehlert}, {Enoto}, {Evangelista}, {Fabiani},
  {Ferrazzoli}, {Garcia}, {Gunji}, {Hayashida}, {Iwakiri}, {Jorstad}, {Karas},
  {Kitaguchi}, {Kolodziejczak}, {Krawczynski}, {Latronico}, {Liodakis},
  {Maldera}, {Manfreda}, {Marin}, {Marinucci}, {Marscher}, {Matt}, {Mitsuishi},
  {Mizuno}, {Ng}, {O'Dell}, {Omodei}, {Oppedisano}, {Papitto}, {Pavlov},
  {Peirson}, {Perri}, {Pesce-Rollins}, {Petrucci}, {Pilia}, {Possenti},
  {Puccetti}, {Ramsey}, {Rankin}, {Ratheesh}, {Romani}, {Sgr{\`o}}, {Slane},
  {Soffitta}, {Spandre}, {Tamagawa}, {Tavecchio}, {Taverna}, {Tawara},
  {Tennant}, {Thomas}, {Tombesi}, {Trois}, {Vink}, {Weisskopf}, {Wu}, {Xie},
  {Zane}, \& {IXPE Collaboration}}]{Tsygankov2022}
{Tsygankov}, S.~S., {Doroshenko}, V., {Poutanen}, J., {et~al.} 2022, \apjl,
  941, L14

\bibitem[{{Veledina} {et~al.}(2023){Veledina}, {Muleri}, {Poutanen},
  {Podgorn{\'y}}, {Dov{\v{c}}iak}, {Capitanio}, {Churazov}, {De Rosa}, {Di
  Marco}, {Forsblom}, {Kaaret}, {Krawczynski}, {La Monaca}, {Loktev},
  {Lutovinov}, {Molkov}, {Mushtukov}, {Ratheesh}, {Rodriguez Cavero},
  {Steiner}, {Sunyaev}, {Tsygankov}, {Zdziarski}, {Bianchi}, {Bright},
  {Bursov}, {Costa}, {Egron}, {Garcia}, {Green}, {Gurwell}, {Ingram}, {Kajava},
  {Kale}, {Kraus}, {Malyshev}, {Marin}, {Matt}, {McCollough}, {Mereminskiy},
  {Nizhelsky}, {Piano}, {Pilia}, {Pittori}, {Rao}, {Righini}, {Soffitta},
  {Shevchenko}, {Svoboda}, {Tombesi}, {Trushkin}, {Tsybulev}, {Ursini},
  {Weisskopf}, {Wu}, {Agudo}, {Antonelli}, {Bachetti}, {Baldini},
  {Baumgartner}, {Bellazzini}, {Bongiorno}, {Bonino}, {Brez}, {Bucciantini},
  {Castellano}, {Cavazzuti}, {Chen}, {Ciprini}, {Del Monte}, {Di Gesu}, {Di
  Lalla}, {Donnarumma}, {Doroshenko}, {Ehlert}, {Enoto}, {Evangelista},
  {Fabiani}, {Ferrazzoli}, {Gunji}, {Hayashida}, {Heyl}, {Iwakiri}, {Jorstad},
  {Karas}, {Kislat}, {Kitaguchi}, {Kolodziejczak}, {Latronico}, {Liodakis},
  {Maldera}, {Manfreda}, {Marinucci}, {Marscher}, {Marshall}, {Massaro},
  {Mitsuishi}, {Mizuno}, {Negro}, {Ng}, {O'Dell}, {Omodei}, {Oppedisano},
  {Papitto}, {Pavlov}, {Peirson}, {Perri}, {Pesce-Rollins}, {Petrucci},
  {Possenti}, {Puccetti}, {Ramsey}, {Rankin}, {Roberts}, {Romani}, {Sgr{\`o}},
  {Slane}, {Spandre}, {Swartz}, {Tamagawa}, {Tavecchio}, {Taverna}, {Tawara},
  {Tennant}, {Thomas}, {Trois}, {Turolla}, {Vink}, {Xie}, \&
  {Zane}}]{Veledina23}
{Veledina}, A., {Muleri}, F., {Poutanen}, J., {et~al.} 2023, Nature Astronomy,
  in press, arXiv:2303.01174

\bibitem[{{Weisskopf} {et~al.}(2022){Weisskopf}, {Soffitta}, {Baldini},
  {Ramsey}, {O'Dell}, {Romani}, {Matt}, {Deininger}, {Baumgartner},
  {Bellazzini}, {Costa}, {Kolodziejczak}, {Latronico}, {Marshall}, {Muleri},
  {Bongiorno}, {Tennant}, {Bucciantini}, {Dovciak}, {Marin}, {Marscher},
  {Poutanen}, {Slane}, {Turolla}, {Kalinowski}, {Di Marco}, {Fabiani},
  {Minuti}, {La Monaca}, {Pinchera}, {Rankin}, {Sgro'}, {Trois}, {Xie},
  {Alexander}, {Allen}, {Amici}, {Andersen}, {Antonelli}, {Antoniak},
  {Attin{\`a}}, {Barbanera}, {Bachetti}, {Baggett}, {Bladt}, {Brez}, {Bonino},
  {Boree}, {Borotto}, {Breeding}, {Brienza}, {Bygott}, {Caporale}, {Cardelli},
  {Carpentiero}, {Castellano}, {Castronuovo}, {Cavalli}, {Cavazzuti},
  {Ceccanti}, {Centrone}, {Citraro}, {D'Amico}, {D'Alba}, {Di Gesu}, {Del
  Monte}, {Dietz}, {Di Lalla}, {Persio}, {Dolan}, {Donnarumma}, {Evangelista},
  {Ferrant}, {Ferrazzoli}, {Ferrie}, {Footdale}, {Forsyth}, {Foster},
  {Garelick}, {Gunji}, {Gurnee}, {Head}, {Hibbard}, {Johnson}, {Kelly},
  {Kilaru}, {Lefevre}, {Roy}, {Loffredo}, {Lorenzi}, {Lucchesi}, {Maddox},
  {Magazzu}, {Maldera}, {Manfreda}, {Mangraviti}, {Marengo}, {Marrocchesi},
  {Massaro}, {Mauger}, {McCracken}, {McEachen}, {Mize}, {Mereu}, {Mitchell},
  {Mitsuishi}, {Morbidini}, {Mosti}, {Nasimi}, {Negri}, {Negro}, {Nguyen},
  {Nitschke}, {Nuti}, {Onizuka}, {Oppedisano}, {Orsini}, {Osborne}, {Pacheco},
  {Paggi}, {Painter}, {Pavelitz}, {Pentz}, {Piazzolla}, {Perri},
  {Pesce-Rollins}, {Peterson}, {Pilia}, {Profeti}, {Puccetti}, {Ranganathan},
  {Ratheesh}, {Reedy}, {Root}, {Rubini}, {Ruswick}, {Sanchez}, {Sarra},
  {Santoli}, {Scalise}, {Sciortino}, {Schroeder}, {Seek}, {Sosdian}, {Spandre},
  {Speegle}, {Tamagawa}, {Tardiola}, {Tobia}, {Thomas}, {Valerie}, {Vimercati},
  {Walden}, {Weddendorf}, {Wedmore}, {Welch}, {Zanetti}, \& {Zanetti}}]{IXPE22}
{Weisskopf}, M.~C., {Soffitta}, P., {Baldini}, L., {et~al.} 2022, JATIS, 8,
  026002

\end{thebibliography}


\end{document}